\documentclass[12pt,preprint]{aastex}

\usepackage{amsmath}
\usepackage{appendix}
\usepackage{subfigure}
\usepackage{booktabs}
\usepackage{graphicx}
\usepackage{rotating}
\begin{document}

\title{Evolution of LMXBs under Different Magnetic Braking Prescriptions
}
\author{
Zhu-Ling Deng$^{1,2,3,4,5}$, Xiang-Dong Li$^{4,5*}$, Zhi-Fu Gao$^{1,2*}$, Yong Shao$^{4,5}$}

\affil{$^{1}$Xinjiang Astronomical Observatory, Chinese Academy of Sciences, 150, Science 1-Street, Urumqi, Xinjiang 830011, China}

\affil{$^{2}$Key Laboratory of Radio Astronomy, Chinese Academy of Sciences, Nanjing 210008, China}

\affil{$^{3}$University of Chinese Academy of Sciences, 19A Yuquan Road, Beijing 100049, China; zhifugao@xao.ac.cn}

\affil{$^{4}$School of Astronomy and Space Science, Nanjing University, Nanjing 210023, China; lixd@nju.edu.cn}

\affil{$^{5}$Key Laboratory of Modern Astronomy and Astrophysics (Nanjing University), Ministry of
Education, Nanjing 210023, China}

\begin{abstract}
Magnetic braking (MB) likely plays a vital role in the evolution of low-mass X-ray binaries (LMXBs). However, it is still uncertain about the physics of MB, and there are various proposed scenarios for MB in the literature. To examine and discriminate the efficiency of MB, we investigate the LMXB evolution with five proposed MB laws. Combining detailed binary evolution calculation with binary population synthesis, we obtain the expected properties of LMXBs and their descendants binary millisecond pulsars. We then discuss the strength and weakness of each MB law by comparing  the calculated results with observations. We conclude that the $\tau$-boosted MB law seems to best match the observational characteristics.

\end{abstract}


\section{INTRODUCTION}
Low-mass X-ray binaries (LMXBs) contain an accreting compact star (a black hole or a neutron star) and a low-mass donor. Mass transfer (MT) in LMXBs proceeds via Roche-lobe overflow (RLOF). There are about 200 LMXBs discovered in the Galaxy \citep{Liu07}, and their formation remains to be a controversial topic \citep[][for reviews]{T06,Li15}. Here, we focus on the evolution of LMXBs with a neutron star (NS).
%It has been proposed that a large fraction of LMXBs have evolved from intermediate-mass X-ray binaries (IMXBs), which undergo rapid mass transfer on a (sub)thermal timescale. That will cause the donor star to lose most its mass and become a low-mass star \citep{Ko00,Po00,P03,T00,SL12,De20}.
For these LMXBs there exists a critical value  for the initial orbital period, i.e., the bifurcation period $P_{\rm bif}$, separating the formation of converging LMXBs from diverging LMXBs \citep{Py88,Py89}. The origin of the bifurcation period comes from the competition between orbital expansion caused by MT and orbital shrinking due to angular momentum loss (AML) by magnetic braking (MB) and gravitational radiation (GR), both of which strongly depend on the orbital period. Binaries with $P_{\rm orb}>P_{\rm bif}$ will become relatively wide binaries containing a recycled NS and a He or CO white dwarf (WD), while those with $P_{\rm orb}<P_{\rm bif}$ mainly keep going to contract along the cataclysmic variable (CV)-like or ultra-compact X-ray binary (UCXB) evolutionary tracks \citep{Ra95,KW96,KR99,TS99,Ko00,Po00,T00,Li02,P02,Lin11,I14a,JL14}. Theoretically, the value of the bifurcation period $P_{\rm bif}$ is $\sim 0.5-1$ d, but it sensitively depends on the MB models \citep{P02,ML09,Va19}

AML can greatly influence the orbital separation and the MT rate, thus playing a significant role in the evolution of LMXBs \citep{ML09,I14a,PI16,Va19,VI19}. Though GR and MB are both the main AML mechanisms in LMXBs, GR dominates in systems with orbital periods shorter than a few hours, and has been well understood theoretically and confirmed by observations of gravitational wave events \citep{LV16,LV17,WH16}. For LMXBs with longer orbital periods, the dominant mechanism of AML is  MB. Note that the idea of MB was originally motivated by the studies of spin evolution of Solar-type stars \citep{B91}. The main idea is that AM is carried away by the stellar wind, which is forced by the magnetic fields to co-rotate with the star  out to a certain distance (5 to 10 stellar radii or more).

The \citet{Sk72} MB prescription is most popular in the simulations of the evolution of low-mass stars. Based on the Skumanich MB model, \citet{VZ81} and \citet{Ra83} suggested an empirical MB formula for LMXBs, which has been widely used in the evolutionary investigations. However, it has been pointed out that the mass accretion rates inferred from observations are higher than the calculated MT rates based on that MB law by about an order of magnitude \citep{P02,P03,SL15,PI16,Va19}. Moreover, the predicted orbital period ($P_{\rm orb}$) distribution of binary millisecond pulsars (BMSPs), which are thought to be the descendants of LMXBs, does not match the observations, especial for BMSPs with $P_{\rm orb}\sim 0.1-10$ d \citep{P03,I14a,SL15}. It was also found that UCXBs can hardly form within a Hubble time under the Skumanich MB law \citep{P02,van05a}.

Some alternative AML mechanisms have been proposed to account for the discrepancy between theory and observation. For example, \citet{CL06} proposed that a circumbinary disk can extract orbital AM from the inner LMXB.  \citet{Ju06} suggested that the secondaries of black hole intermediate-mass X-ray binaries (BHIMXBs) may possess anomalously high magnetic fields to enhance the efficiency of MB. The effect of consequential AML was also investigated by \citet{Ne16} and \citet{Sc16}  for the CV evolution. \citet{PI16} reproduced the formation of Sco X-1 by using a modified Skumanich MB law. Recently, \citet{Va19} added two additional ``boosted" factors (i.e., convective turnover time and stellar wind) to the Skumanich MB, in order to explain the measured MT rates of LMXBs. \citet[][hereafter VI19]{VI19} suggested a new MB formula (named as Convection And Rotation Boosted MB) for persistent NS LMXBs.

Meanwhile, various MB models have been developed to investigate the rotational evolution of single low-mass stars \citep[e.g.,][]{Ma12,RM12,GB13,VS13,SA17}. In principle, these MB mechanisms may also work in LMXBs. So it is important to examine and discriminate different MB formulae by comparing the observations of LMXBs and BMSPs with the theoretical results, and this is the objective of our study.

The structure of the paper is organized as follows. We describe the stellar evolution code and summarize several typical MB formalisms in Section 2. The calculated binary evolution results are demonstrated in Section 3. We present our discussion and summary in Section 4.

\section{EVOLUTION CODE AND BINARY MODEL}

\subsection{The stellar evolution code}
All calculations were carried out by using the stellar evolution code Modules for Experiments in Stellar Astrophysics (MESA; version number 11554; Paxton et al. 2011, 2013, 2015, 2018, 2019). The binaries initially consist of an NS (of mass $M_1$) and a main-sequence (MS) secondary/donor star (of mass $M_2$) with Solar chemical compositions ($X=0.7$ and $Z=0.02$). When treating convection we adopt the mixing length parameter $\alpha =2$, and do not consider semiconvection and overshooting. The effective radius of the RL for the secondary is calculated with the \citet{E83} formula,

\begin{equation}
\frac{R_{\rm L,2}}{a}=\frac{0.49 q^{2 / 3}}{0.6 q^{2 / 3}+\ln \left(1+q^{1 / 3}\right)},
\end{equation}
where $a$ is the orbital separation of the binary and $q=M_{2}/M_{1}$ is the mass ratio. We adopt the \citet{R88} scheme to calculate the MT rate via RLOF,
\begin{equation}
-\dot{M}_{2}=\dot{M}_{2,0} \exp \left(-\frac{R_{2}-R_{\mathrm{L}, 2}}{H}\right),
\end{equation}
where $H$ is the scale-height of the atmosphere evaluated at the surface of the donor, $R_{2}$ is the radius of the donor, and
\begin{equation}
\dot{M}_{2,0}=\frac{1}{e^{1 / 2}} \rho c_{\mathrm{th}} Q,
\end{equation}
where $\rho$ and $c_{\mathrm{th}}$ are the mass density and the sound speed on the surface of the star respectively, and $Q$ is the cross section of the mass flow via the $L_{1}$ point.

\subsection{Mass and angular momentum loss}

In LMXBs the binary orbital revolution and the spin of the secondary star are assumed to be synchronized, because the tidal synchronization timescale is usually much shorter than the evolutionary timescale of LMXBs \citep{K88}\footnote{We actually treated the spin and orbital evolution separately in our calculation, since in a few cases the tidal torques are not strong enough to synchronize the spin and orbital revolution.}. The orbital AM can be expressed as
\begin{equation}
J_{\mathrm{orb}}=\frac{M_{1} M_{2}}{M_{1}+M_{2}} \Omega a^{2},
\end{equation}
where the orbital angular velocity $\Omega=\sqrt{GM/a^{3}}$, and $M=M_1+M_2$ is the total mass. Taking the logarithmic derivative of Eq.~(4) with respect to time gives the rate of change in the orbital separation
\begin{equation}
\frac{\dot{a}}{a}=2 \frac{\dot{J}_{\mathrm{orb}}}{J_{\mathrm{orb}}}-2 \frac{\dot{M}_{1}}{M_{1}}-2 \frac{\dot{M}_{2}}{M_{2}}+\frac{\dot{M}}{M}.
\end{equation}
Here the total rate of change in the orbital AM is determined by
\begin{equation}
\dot{J}_{\rm orb}=\dot{J}_{\rm gr}+\dot{J}_{\rm ml}+\dot{J}_{\rm mb},
\end{equation}
where the three terms on the right-hand-side of Eq.~(6) represent AML caused by GR, mass loss, and mb, respectively. The GR-induced AML rate $\dot{J}_{\rm gr}$ is given by \citep{L59,F71}
\begin{equation}
\frac{\dot{J}_{\mathrm{gr}}}{J_{\mathrm{orb}}}=-\frac{32 G^{3}}{5 c^{5}} \frac{M_{1} M_{2} M}{a^{4}},
\end{equation}
where $G$ and $c$ are the gravitational constant and the speed of light, respectively. The accretion rate of the NS is assumed to be limited by the Eddington accretion rate $\dot{M}_{\mathrm{Edd}}$,
\begin{equation}
\dot{M}_{\mathrm{1}}=\min \left(\left|\dot{M}_{2}\right|, \dot{M}_{\mathrm{Edd}}\right),
\end{equation}
and the mass loss rate from the binary system is
\begin{equation}
\dot{M}=\dot{M}_{1}-\left|\dot{M}_{2}\right|.
\end{equation}
In the case of super-Eddington MT, we adopt the isotropic reemission model, assuming that the extra material leaves the binary in the form of isotropic wind from the NS. Therefore, the AML rate due to mass loss is
\begin{equation}
\dot{J}_{\mathrm{ml}}=-\left(\left|\dot{M}_{2}\right|-\dot{M}_{\mathrm{1}}\right) a_{\mathrm{1}}^{2} \Omega,
\end{equation}
where $a_{\mathrm{1}}$ is the distance between the NS and the center of mass of the binary.

Magnetized stellar winds coupled with the secondary star can reduce the stellar spin speed effectively, then carry away the orbital AM through tidal torques.
In the following, we list several MB prescriptions we will use in calculating the evolution of LMXBs.

1. The Skumanich Model

Based on the widely used \citet{Sk72} MB model, \citet{VZ81} and \citet{Ra83} proposed an MB formula for LMXBs:
\begin{equation}
\dot{J}_{\mathrm{mb,Sk}}=-3.8 \times 10^{-30} M_{2} R_{\odot}^4 \left(\frac{R_2}{R_{\odot}}\right)^{\gamma_{\rm mb}}  \Omega^{3} \  \rm dyne \,cm.
\end{equation}
Here $R_{\odot}$ is the Solar radius and $\gamma_{\rm mb}$ is a dimensionless parameter ranging from 0 to 4. In this paper we adopt the default value $\gamma_{\rm mb}=4$ in MESA. Note that for  CV evolution smaller values of $\gamma_{\rm mb}$ are often used \citep{Kn11}, so we also calculate the evolution of LMXB with $\gamma_{\rm mb} =2$ and 3, and find that the main results do not change significantly compared with those with $\gamma_{\rm mb} =4$.

2. The Matt12 Model

\citet[][hereafter MP08]{Ma08} carried out 2D axisymmetric magnetohydrodynamical (MHD) simulations of wind outflows from a rigidly rotating star with a aligned dipolar field, to determine the dependence of the wind torque on the magnetic field and mass outflow rate. \citet[][hereafter Matt12]{Ma12} extended the parameter study of MP08 by including variations in both the magnetic field strength and the stellar rotation rate, and derived the following stellar wind torque formula,
\begin{equation}
\frac{\mathrm{d} J}{\mathrm{d} t}=-\frac{K_{1}^{2}}{(2 G)^{\rm m}} \bar{B}_{\rm s}^{\rm 4 m} \dot{M}_{\rm 2,W}^{\rm 1-2m} \left(\frac{R_2^{\rm 5 m+2}}{M_2^{\rm m}}\right) \frac{\Omega}{\left(K_{2}^{2}+0.5 u^{2}\right)^{\rm m}},
\end{equation}
where $u$ is the equatorial rotation speed divided by the break-up speed, $K_1=6.7$, $K_2=0.506$, and $m=0.17$ are adjustable parameters to fit the observations, which were taken from \citet{GB13}. The wind loss rate $\dot{M}_{\rm 2,W}$ is evaluated using the \citet{Re75} wind mass-loss prescription
\begin{equation}
\dot{M}_{\rm 2,W}=4\times 10^{-13}\,M_{\odot}{\rm yr}^{-1} \left(\frac{R_2}{R_{\odot}}\right) \left(\frac{L_2}{L_{\odot}}\right) \left(\frac{M_{\odot}}{M_2}\right).
\end{equation}
where $L$ is the donor's luminosity and $L_{\odot}$ is the Solar luminosity.
The mean magnetic field $\bar{B}_{\rm s}=fB_{\rm s}$, where $B_{\rm s}$ is the surface magnetic field strength of the donor star, and $f$ is the filling factor expressing the magnetized fraction of the stellar surface \citep{Sa96,Am16}. Here we set $B_{\rm s}$ as the Solar surface magnetic field strength $B_{\rm s}=B_{\odot}=1$ G. The expression of $f$ is re-calibrated to reach the Solar mass-loss value at the age of the Sun for solid-body rotating models,
\begin{equation}
f=\frac{0.4}{[1+(x/0.16)^{2.3}]^{1.22}},
\end{equation}
with $x$ being the normalised Rossby number $x=(\frac{\Omega_{\odot}}{\Omega})(\frac{\tau_{\odot,\rm conv}}{\tau_{\rm conv}})$ \citep{Am16}. Here the rotation rate of the Sun $\Omega_{\odot}\simeq 3\times 10^{-6}$ $\mathrm{s}^{-1}$, $\tau_{\mathrm{conv}}$ is the turnover time of the convective eddies of the donor star \citep{No84}, and $\tau_{\odot, \rm conv}\simeq 2.8\times 10^6$ s.

3. The RM12 Model

\citet[][hereafter RM12]{RM12} proposed a formalism in which MB is related to mean surface magnetic field strength ($fB_{\rm s}\propto \Omega^a$) instead of magnetic flux ($B_{\rm s}R^{2}\propto \Omega^a$) as suggested by \citet{K88}. In this model there is a critical surface angular velocity $\Omega_{\mathrm{sat}}=3\Omega_{\odot}$, above which the magnetic field reaches saturation \citep{SH87}. This saturation modifies the braking law and the relation between the AML rate and the rotation rate, and the magnetic field strength likely stops increasing even if the star still spins up \citep{Vi84,O95}. The AML rate is given by
\begin{equation}
\frac{\mathrm{d} J}{\mathrm{d} t}=-C\left(\frac{R^{16}}{M^{2}}\right)^{1 / 3} \Omega_{\rm s} \quad \text { for } \Omega \geq \Omega_{\mathrm{sat}},
\end{equation}
\begin{equation}
\frac{\mathrm{d} J}{\mathrm{d} t}=-C\left(\frac{R^{16}}{M^{2}}\right)^{1 / 3}\left(\frac{\Omega_{\rm s}}{\Omega_{\mathrm{sat}}}\right)^{4} \Omega\quad \text { for } \Omega<\Omega_{\mathrm{sat}},
\end{equation}
%\textbf{where $\Omega_{\rm s}$ is the secondary star spin speed, and we adopt the initial spin speed $\Omega_{\rm s}=7.27\times 10^{-5}$ (with spin period $P_{s}=1$ d), and}
with
\begin{equation}
C=\frac{2}{3}\left(\frac{B_{\text {crit }}^{8}}{G^{2} K_{V}^{4} \dot{M}_{\rm 2, W}}\right)^{1 / 3}.
\end{equation}
Here \citet{RM12} assumed that the saturation field strength $B_{\rm crit }$, the wind mass-loss rate $\dot{M}_{\rm 2, W}$ and the velocity scaling factor $K_V$ are all constant, independent of stellar mass. By comparing with observations they got a best-fit choice with $C=2.66\times 10^3\, \rm (g^5 cm^{-10} s^3)^{1/3}$.
%\textbf{Note that we find that the tidal synchronization timescale between the binary orbital revolution and the spin of the secondary star is not always much shorter than the evolutionary timescale of LMXBs under this MB law. To ensure our results reliably, when using RM12 MB law, we adopt $\Omega$ and $\Omega_{\rm s}$ to represent the binary orbital angular velocity and the spin speed of the secondary star, respectively. The tidal torque equation be adopted in \citet{Hut81}. Tidal torque becomes a medium to extract binary orbital AM from spin AM of secondary star.}

%being a constant ($C=10^{39}$) because the parameters of the right-hand side are assumed to adjust to keep the overall product constant throughout the evolution.

%This model is also good agreement with the empirical Skumanich law. While Reiners and Mohanty suggested an error of the MB formula of \citet[][hereafter K88]{K88}, and the error arises due to a confusion between magnetic field strength and magnetic flux (adopted $B_{\rm s}R^2\propto \Omega^a$ instead of $fB_{\rm s}\propto \Omega^a$). The error result in the formulation of K88 that a fundamental dependence on radius has been missed \citep{RM12}.

4. The $\tau$-boosted Model

\citet{Va19} suggested that the surface magnetic field $B_{\rm s}$ of the secondary is connected with the Rossby number $R_0$ as \citep{No84,I06},
\begin{equation}
\frac{B_{\rm s}}{B_{\rm s,\odot}}=\frac{R_0}{R_{0,\odot}}=\left(\frac{\Omega}{\Omega_{\odot}}\right)\left(\frac{\tau_{\rm conv}}{\tau_{\odot, \rm conv}}\right).
\end{equation}
In addition, they added winds in the Skumanich MB and proposed a modified MB law:
\begin{equation}
\dot{J}_{\mathrm{MB},\text{boost}}=\dot{J}_{\mathrm{MB},\mathrm{Sk}}\left(\frac{\Omega}{\Omega_{\odot}}\right)^{\beta}\left(\frac{\tau_{\text{conv}}}{\tau_{\odot, \rm conv}}\right)^{\xi}\left(\frac{\dot{M}_{\mathrm{2,W}}}{\dot{M}_{\odot }}\right)^{\alpha}.
\end{equation}
This MB law can be divided into three forms depending on the power indices ($\xi$, $\alpha$, $\beta$). They are classified into Convection-boosted ($\tau$-boosted, with $\xi=2$, $\alpha=0$, $\beta=0$), Intermediate (with $\xi=2$, $\alpha=1$, $\beta=0$) and Wind-boosted (with $\xi=4$, $\alpha=1$, $\beta=2$), respectively. The $\tau$-boosted and Intermediate MB schemes were found to be more effective to reproduce the observed NS LMXBs \citep{Va19}. Here, we adopt the $\tau$-boosted scheme.

5. The VI19 Model

The Intermediate prescription in \citet{Va19} was found to have difficulty in explaining the effective temperature of Sco X-1. \citet{VI19} then considered the rotational effects on the Alfv\'en radius and the magnetic field dependence on the convective turnover time, and presented a modified MB prescription called Convection And Rotation Boosted (CARB),
\begin{equation}
\begin{aligned}
\dot{J}_{\mathrm{MB, CARB}}=&-\frac{2}{3} \dot{M}_{\mathrm{2,W}}^{-1 / 3} R^{14 / 3}\left(v_{\mathrm{esc}}^{2}+2 \Omega^{2} R^{2} / K_{2}^{2}\right)^{-2 / 3}  \Omega_{\odot} B_{\odot}^{8 / 3}\left(\frac{\Omega}{\Omega_{\odot}}\right)^{11/3}\left(\frac{\tau_{\mathrm{conv}}}{\tau_{\odot, \rm conv}}\right)^{8 / 3},
\end{aligned}
\end{equation}
where $v_{\rm esc}$ is the surface escape velocity, and $K_2=0.07$ is a constant obtained from a grid of simulations by \citet{Re15}.

Finally, by default, we assume that MB operates only when the star has a convective envelope and a radiative core \citep{P15}.

\section{RESULTS OF EVOLUTION CALCULATIONS}
In our calculations, we choose the initial NS mass $M_1=1.3M_{\odot}$ and the donor's mass ranging from $M_2=1.0M_{\odot}$ to $M_2=4.0M_{\odot}$. We adopt the initial binary orbital period in the range $-0.5\leq \mathrm{log}(P_{\rm orb,i}/\text{d})\leq 2$ in steps of $\Delta \mathrm{log}(P_{\rm orb,i}/\text{d})=0.05$.

\subsection{Example Evolutions with Different MB Laws}
We first demonstrate the evolutionary sequences for an LMXB that consists of a $1.1M_{\odot}$ donor star with different MB laws in Fig.~1. The left, middle, and right panels correspond to the initial orbital period $P_{\rm orb,i}=$ 1.0, 10.0, and 100.0 d, respectively. In the upper panels we depict the evolution of the orbital period as a function of the donor mass, and in the lower panels we compare the the AML rates for different MB laws. MB stops working when the donor star becomes full convective, with mass $\sim 0.1-0.3M_{\odot}$. Note that there is a jump in the AML rate for the Matt12 MB at the final evolutionary stage with $P_{\rm orb}\simeq 1$ d. The reason is the donor star has evolved into a (proto-)He WD, and the residual shell hydrogen burning and vigorous flashes form a convective hydrogen envelope and cause repeated RLOF \citep{I14b}.

The upper panels of Fig.~1 show that different MB laws lead to diverse evolutionary paths with the same initial parameters. When $P_{\rm orb,i}= 1.0$ d, the orbital period increases with the Matt12 MB law, but decreases in all other cases. When $P_{\rm orb,i}$ is set to be 10 d, binaries in the Skumanich MB model join the divergent sequences. In contrast, the orbital period still decreases with the RM12 MB law. In the other two cases with the $\tau$-boosted and VI19 MB, the final period does not significantly deviate from its initial value. When $P_{\rm orb,i}$ is increased to 100 d, the binaries evolve to divergent systems in all cases. The lower panels of Fig.~1 display the AML rate due to MB relative to the total AML rate. Generally, for the same initial parameters, stronger MB leads to forming converging systems, while weaker MB law result in forming divergent binary pulsar systems.

Table 1 presents the final parameters of the evolutionary products, including the NS mass, companion mass, orbital period and hydrogen abundance. For the LMXB with $P_{\rm orb,i}=1$ d, only the Matt12 MB leads to the formation of a binary pulsar  with a $0.25M_{\odot}$ WD (with the final orbital period $P_{\rm orb, f}=13.367$ d), while in other cases the binaries evolve to converging binaries ($P_{\rm orb, f} < 0.3$ d) that contain a low-mass and hydrogen-rich companion star. The final NSs are heaviest and lightest under the Skumanich and the VI19 MB laws, respectively. For the LMXB systems with $P_{\rm orb,i}=10.0$ d, using the Skumanich and the Matt12 MB laws form wide binary pulsars, and using the other MB laws form tight binaries. In the latter cases the MT rate can become super-Eddington which significantly lowers the accretion efficiency of the NSs. All LMXB systems with $P_{\rm orb,i}=100$ d form divergent wide binary pulsars with a WD companion.

%The Fig. 12 shows the more specific evolution of different MB laws. The LMXB consist of a NS and a donor star with initial masses $M_{1}=1.3M_{\odot}$ and $M_2=1.4M_{\odot}$ respectively, and initial orbital $P_{\rm orb}=1.0$ day. Different MB schemes result in different evolutionary tracks for the same initial conditions. For the `Matt12' MB law, final product is a WD BMSP system,  and the others MB laws would form converging systems.

\subsection{Comparing the Properties of LMXBs with Observations}

We compare the properties of LMXBs from our evolution calculations with observations. In Fig.~2-6 we present the evolution of the orbital period as a function of the donor mass with different MB laws. In each figure we consider initial donor mass to be 1.0, 1.1, 1.2, 1.4, 2.0, and 3.0 $M_{\odot}$. We use different colors to represent the ranges of the MT rates. In IMXBs, MT proceeds on a (sub)thermal timescale because of the relatively high mass ratio ($q=M_2/M_1>1.5$), but the stability depends on both the mass ratio and the evolutionary state of the donor star \citep{Pa76,We84,IL93,T00,Lin11,T11,SL12,JL14,De20}. We stop calculation and regard the MT to be unstable when the MT rate becomes $>10^{-4} M_{\odot} {\rm yr}^{-1}$. We adopt the observational parameters from Tables 4 and 5 of \citet{Va19} for the binned properties of selected persistent and transient NS LMXBs. Note that some systems are in globular clusters (GCs) and probably formed via dynamical encounters rather than from primordial binaries \citep{Va19}. We use circles, squares and triangles to represent persistent, GC, and transient LMXBs\footnote{For transient LMXBs we adopt long-term, averaged MT rates inferred from observations for comparison.}, respectively.

Fig.~2 compares the calculated results with the Skumanich MB law with observations. It shows that most LMXBs can be covered in the $P_{\rm orb}-M_{2}$ diagram, but the accretion rates in persistent NS LMXBs are at least an order of magnitude higher than the theoretically expected ones \citep[see also][]{P02,SL15,Va19}. Meanwhile, the mean accretion rates in part of the transient NS LMXB systems are lower than the calculations, and there are very few systems that can form UCXBs.

Fig.~3 shows the results with the $\tau$-boosted MB law. With this MB law most LMXBs can be reproduced. In addition, the calculated rates can match the observed accretion rates for almost all LMXBs. UCXBs can be produced more effectively compared with the case using the Skumanish MB.

Fig.~4 shows the results with the Matt12 MB law. Its strength of AML is so weak that it hardly reproduces converging systems with $P_{\rm orb}<1$ d. For relatively compact LMXBs with specific orbital period and donor mass, their calculated MT rates are lower than inferred from observations by about two orders of magnitude.

Fig.~5 shows the results with the RM12 MB law. While it can reproduce all LMXBs in the $P_{\rm orb}-M_{2}$ diagram, the calculated MT rates for most LMXBs do not match the observations, but UCXBs can form effectively.

Fig.~6 shows the results with the VI19 MB law. The evolutionary tracks can cover all LMXBs in the $P_{\rm orb}-M_{2}$ diagram. The calculated MT rates can match the observations of persistent LMXBs, but are too high for some transient systems.

To summarize the results in Figs.~2-6 in a more clear way, we compare the effectiveness of the five MB models in explaining the LMXBs in Table 2. The symbols $\triangle$, $\blacktriangle$, $\blacktriangle \blacktriangle$ and $\blacktriangle \blacktriangle \blacktriangle$ represent that the LMXB can be reproduced by 0, $1-2$, $3-5$, and more than 5 evolutionary tracks under a specific MB model, respectively. It seems that both the $\tau$-boosted and the VI19 MB laws are more preferred.

\subsection{Comparing the Orbital Period Distribution of Binary Pulsars with observations}

A large fraction of LMXBs will evolve into BMSPs by the ``recycling" process \citep{A82}. So a successful MB model should be able to reproduce the properties of not only LMXBs but also their descendants BMSPs.
\citet{SL15} used the binary population synthesis (BPS) code developed by \citet{Hu02} to investigate the formation and evolution of Galactic intermediate- and low-mass X-ray binaries (I/LMXBs). Some key assumptions of the initial parameters in that study are briefly described as follows. A constant star formation rate of $5M_{\odot}$ yr$^{-1}$ in the Galaxy was assumed for the primordial binaries with Solar metallicity \citep{Sm78}.
The primary star's mass is distributed according to the initial mass function of \citet{Kr93} in the range of $3-30 M_{\odot}$, and the initial mass ratio of the secondary to the primary masses has a uniform distribution between 0 and 1.
If the mass transfer is dynamically unstable, a common envelope (CE) phase follows. The standard energy conservation equation \citep{We84} was used to treat the CE evolution with an energy efficiency $\alpha_{\rm CE}=1.0$ and the binding energy parameters $\lambda$ of the primary's envelope taken from the numerical results in \citet{XL10}. NSs formed from core-collapse and electron-capture SNe were assumed to be imparted a kick with a Maxwellian distribution for the kick velocity with $\sigma = 265 $ km\,s$^{-1}$ and 50 km\,s$^{-1}$, respectively \citep{Ho05,De06}.

Combining the birthrate and the distribution of the orbital period and the donor mass for the incipient I/LMXBs in \citet{SL15} with our detailed binary evolution calculations, we then obtain the orbital period distributions for BMSPs, which are displayed in Fig.~7 with red line. From top to bottom are the results by using the Skumanich, the $\tau$-boosted, the Matt12, the RM12 and the VI19 MB laws, respectively.

The NSs in LMXBs may be fully or partially recycled depending on the amount of accreted mass $\Delta M_{\rm NS}$. Accretion of $~0.1M_{\odot}$ material is sufficient to spin up an NS's spin period to milliseconds \citep[e.g.,][]{Ta12}. The actual value of $\Delta M_{\rm NS}$ may be smaller by a factor of 2 if considering the effect of the NS magnetic field-accretion disk interaction \citep{De20}. So we use this criterion to distinguish fully and partially recycled pulsars. In the left and right panels of Fig. 7, we show the calculated distributions of binary pulsars with $\Delta M_{\rm NS}>0$ and $\Delta M_{\rm NS}>0.05\, M_{\odot}$, respectively. The observed distributions of of binary pulsars with spin periods $P_{\rm s} \leq 1$ s and $P_{\rm s} \leq 30$ ms are plotted in blue line correspondingly \citep[data are from the ATNF pulsar catalog\footnote{https://www.atnf.csiro.au/research/pulsar/psrcat/},][]{Ma05}.

%For the Fig. 12, from top panels to bottom panels are WD binary pulsars orbital period distribution by using `Skumanich', `Convection boosted', `Matt12', `RM12' and `VI19' MB laws, respectively. The red and blue curves are calculations results and observations, respectively.  , based on a series of analytic formulae of single star evolution \citep{Hu00},

Fig.~7 shows that adopting the Skumanich MB law cannot reproduce binary pulsar systems with $P_{\rm orb} \sim 0.1-1$ day. And there is a mismatch between calculations and observations about the distribution peak of the orbital period. These conflicts have already been mentioned in previous studies \citep{P03,SL15}. We find that the orbital period distribution predicted by using the $\tau$-boosted and the VI19 MB laws best match the observations. The AML efficiency with the Matt12 MB law is too weak to form binary pulsars with $P_{\rm orb}\lesssim 10$ days. The RM12 MB scheme produces a double-peaked orbital period distribution of binary pulsars which is likely related to the saturation of MB when the angular velocity of the donor becomes large enough. While the calculated orbital period distribution obviously deviates from observations, this model can effectively produce tight binary pulsars with $P_{\rm orb}<1$ d.

We use the Kolmogorov-Smirnov (KS) test to assist inference on their consistency to compare the univariate distributions of the calculated and observed quantities \citep{FJ12}. The calculated and measured orbital periods of the cumulative distribution functions (CDFs) are expressed as $F_{\rm n, orb}$ and $F_{\rm m, orb}$, respectively. We measure the maximum distance $D_{\rm nm}= \max|F_{\rm n, orb}-F_{\rm m,orb}|$ by comparing the CDFs. The null hypothesis is rejected at the confidence level $\alpha=0.05$ if $D_{\rm nm}>D_{\rm \alpha, nm}\simeq 1.36 \sqrt{\frac{n m}{n+m}}$, where $n$ and $m$ are the numbers of calculated and observed binary pulsar systems, respectively. Table~3 shows the calculated values of $D_{\rm nm}$ and $D_{\rm \alpha, nm}$ in the case of the five MB laws. When adopting the $\tau$-boosted MB and the VI19 MB laws, $D_{\rm nm}<D_{\rm \alpha, nm}$, indicating that the null hypothesis cannot be rejected, and the two distributions are not significantly different. Using the other three MB laws, $D_{\rm nm}>D_{\rm \alpha, nm}$, allowing rejection of the null hypothesis and suggesting that the differences are significant.

\subsection{Bifurcation Periods}

In this subsection, we investigate the influence of the MB laws on the bifurcation period. Here we adopt the definition of the bifurcation period used by \citet{P02}, which is the orbital period ($P_{\rm rlof}$) when RLOF just begins, and we express this bifurcation period as $P^{\rm bif}_{\rm rlof}$. Fig.~8 shows the results of the bifurcation periods in the five models described in Section 2.2 with solid lines. We also plot the minimum initial period $P_{\rm ZAMS}$ for a lobe-filling ZAMS donor star with the dashed line.

Several features of the bifurcation periods can be seen in Fig.~8. First, when excluding the RM12 MB law, the magnitude of the bifurcation periods $P^{\rm bif}_{\rm rlof}$ is about $0.4 - 1.8$ d with the companion mass ranging from $1.0M_{\odot}$ to $3.6M_{\odot}$. Second, when adopting the Matt12 MB law, there does not exist any bifurcation period with the donor mass $M_2 \sim 1.3-2.6M_{\odot}$, which means that the binary orbit always increases in these cases. Third, when adopting the RM12 MB law, the final orbital period does not significantly change with $P_{\rm rlof}\sim 1 - 10$ days.

\section{DISCUSSION AND CONCLUSIONS}

\subsection{Forming Binary Pulsars with Orbital Periods $0.1$ d $<P_{\rm orb}<1$ d}

Previous studies on the LMXB evolution suggested that it is hard to form BMSPs with orbital periods of $0.1-1$ d under the conventional Skumanich MB law \citep{P03,SL15}. Moreover, \citet{I14a} found that only a very narrow parameter space is allowed for LMXBs to evolve into tight binary pulsars with a low-mass helium WD companion, which is called the fine-tuning problem for the formation of BMSPs with $P_{\rm orb}\sim 2-9$ hours. These results indicate missing (or misunderstood) AML mechanisms in the LMXB evolution. In this subsection, we try to examine the effectiveness of different MB laws in producing tight binary pulsars.

Fig.~9 shows the orbital period evolution for LMXBs by adopting the five MB laws. We consider LMXBs with initially a $1.0 M_{\odot}$ donor star and the range of the initial orbital period being described as in Section 3. There are totally 51 models in each case. We regard the binary systems to be low-mass binary pulsars when the mass transfer rate $\dot{M}_{\rm tr}<10^{-12}M_{\odot}{\rm yr}^{-1}$, the donor mass $M_{2}>0.13M_{\odot}$, and the hydrogen abundance $< 0.1$.

In Fig.~9, the dashed and dotted curves denote the LMXB evolutions leading to BMSPs with orbital periods $9\,{\rm h}<P_{\rm orb}<24\,{\rm h}$ and $2\,{\rm h}<P_{\rm orb}<9\,{\rm h}$, respectively. No model adopting the Skumanich and the Matt12 MB laws can form the binary pulsars with $2\,{\rm h}<P_{\rm orb}<24\,{\rm h}$. Three and eight models using the RM12 MB law form binary pulsars with $2\,{\rm h}<P_{\rm orb}<9\,{\rm h}$ and $9\,{\rm h}<P_{\rm orb}<24\,{\rm h}$, respectively, and the numbers of successive models are correspondingly one and four when using the $\tau$-boosted MB law, and two and eight when using the VI19 MB law. Note that there is one model adopting the VI19 MB law can form binary pulsar with $2\,{\rm h}<P_{\rm orb}<9\,{\rm h}$, but it continues orbital shrinking (due to AML by GR), evolving to UCXBs within a Hubble time, so we add an annotation in the subplot of Fig. 9 rather than plot it with dotted curve. In general, the stronger MB law can form tight binary pulsar binaries more effectively.

In summary, the VI19, RM12 and $\tau$-boosted MB laws are effective in producing BMSPs with orbital periods 0.1 d $<P_{\rm orb}<$ 1 d, while the other two MB laws require much narrower initial parameter distributions in forming these tight binary pulsars.

%Orbital period evolution for LMXB systems with a donor mass of $1.0 M_{\odot}$, a NS mass of $1.3 M_{\odot}$ and adopting `Skumanich' MB laws. Dashed grey curves denote orbital periods with 2 and 9 hours, and grey solid curve is the time that $1.0 M_{\odot}$ single star terminate main sequence. From top panel to bottom panel, colour denotes current mass accretion rate, donor mass and H$^1$ abundance.

\subsection{Summary}

%Previous studies indicate that there are not a few conflicts between theory and observation in the evolution of LMXBs \citep{P02,P03,van05a,I14a,SL15,Va19}. We concentrate on the effect of MB law in the LMXBs evolution, and try to make up these conflicts using different MB laws.

MB plays a vital role in the LMXB evolution, however, its mechanism has not been well understood. The traditional MB law faces difficulties in explaining the observational characteristics of LMXBs and MSPs, such as the discrepancy between the calculated MT rates and the observed mass accretion rates of LMXBs, and the conflict between the calculated and observed orbital period distribution of BMSPs \citep{P02,P03,van05a,I14a,SL15,Va19}. There are various alternative proposals for the MB laws, and we have made a systematic study to examine their influence on the LMXB evolution. We mainly compare the MT rates of LMXBs and the orbital period distribution of binary pulsars with observations. In addition, we explore the formation efficiency of binary pulsars with $P_{\rm orb}\sim 0.1-1$ d under different MB laws. We summarize our results as follows:

1. Although it is widely used in theoretical investigations of the LMXB evolution, the Skumanich MB law is unable to reproduce the observational accretion rates in LMXBs and the orbital period distribution of binary pulsars.

2. The $\tau$-boosted MB law may be more suitable for the LMXBs evolution compared with other models. The observed properties of most NS LMXBs could be reproduced, including the distributions of the orbital periods, donor masses and MT rates, as well as the orbital period distribution of binary pulsars, especially for the binary pulsars with orbital periods $P_{\rm orb}\sim 0.1-1$ d.

3. The strength of the Matt12 MB scheme is so weak that the calculated MT rates are lower than the observation by at least two orders of magnitude, and it is hard to form converging binaries such as UCXBs and tight binary pulsars.

4. By using the RM12 MB law, we find that both the calculated MT rates of LMXBs and the orbital period distribution of binary pulsars do not match the observations. However, it may play a role in explaining some specific sources like some UCXBs and binary pulsars with $P_{\rm orb}<1$ d.

5. The VI19 MB law can readily account for the characteristics of persistent NS LMXBs and binary pulsars, but the calculated MT rates are higher than the observations of some transient LMXBs by at least an order of magnitude.

Finally we caution that a shortcoming of this work is that we did not take into account several effects that may potentially affect the LMXB evolution, such as accretion disk instabilities, X-ray irradiation of the donor, evaporation of the companion star by the pulsar's high-energy radiation, and consequential AML related with, e.g., circumbinary disks. So our conclusions are based on the assumption that these effects are applicable for specific types of systems rather than the entire population. However, it was proposed that consequential AML likely plays a major role in the evolution of CVs \citep[e.g.,][]{Ne16,Sc16}, and might also be important for NS LMXBs. If that is the case, one needs to carefully discriminate the effect of MB both observationally and theoretically before correctly evaluating the efficiency of different MB prescriptions in the LMXB evolution.

%\subfigure{\includegraphics[scale=0.55]{eps/Mc1.0_t-p.eps}}
%\subfigure{\includegraphics[scale=0.55]{eps/colorbar.eps}}
%\caption{Orbital period evolution for LMXB systems with a donor mass of $1.0 M_{\odot}$, a NS mass of $1.3 M_{\odot}$ and adopt different MB laws. Dashed grey curves denote orbital periods with 2 and 9 hours, and grey solid curve is the time that $1.0 M_{\odot}$ single star terminate main sequence.
%   \label{figure5}}

%\end{sidewaysfigure}

%\caption{Comparison of the calculated orbital period distributions of BMSPs with observations, which are shown with the red and blue curves, respectively. From top panels to bottom panels are different results by 'Skumanich', 'tau-boosted', 'VI19', 'RM12' and 'Matt12' MB laws, respectively. The left panel shows the calculated distributions of pulsar with any accreted mass $\Delta M_{NS}>0$ and the observed distribution of of binary pulsars with spin periods $P_s \leq 1$ s. The center panel shows the calculated distributions of pulsars with accreted mass $\Delta M_{NS}>0.1M_{\odot}$ and the observed distribution of binary pulsars with spin periods $P_s \leq 10$ ms. The right panel is same as the center panel, but we exclude the pulsars of calculations that still in the accretion phase at the evolution end point.

\acknowledgements
This work was supported by the National Key Research and Development Program of China (2016YFA0400803), the Natural Science Foundation of China under grant Nos. 11773015, 12041301, 12041304, and Project U1838201 supported by NSFC and CAS.

%\end{acknowledgements}

\clearpage

\begin{figure}

\centering

\centerline{\includegraphics[angle=0,width=0.8\textwidth]{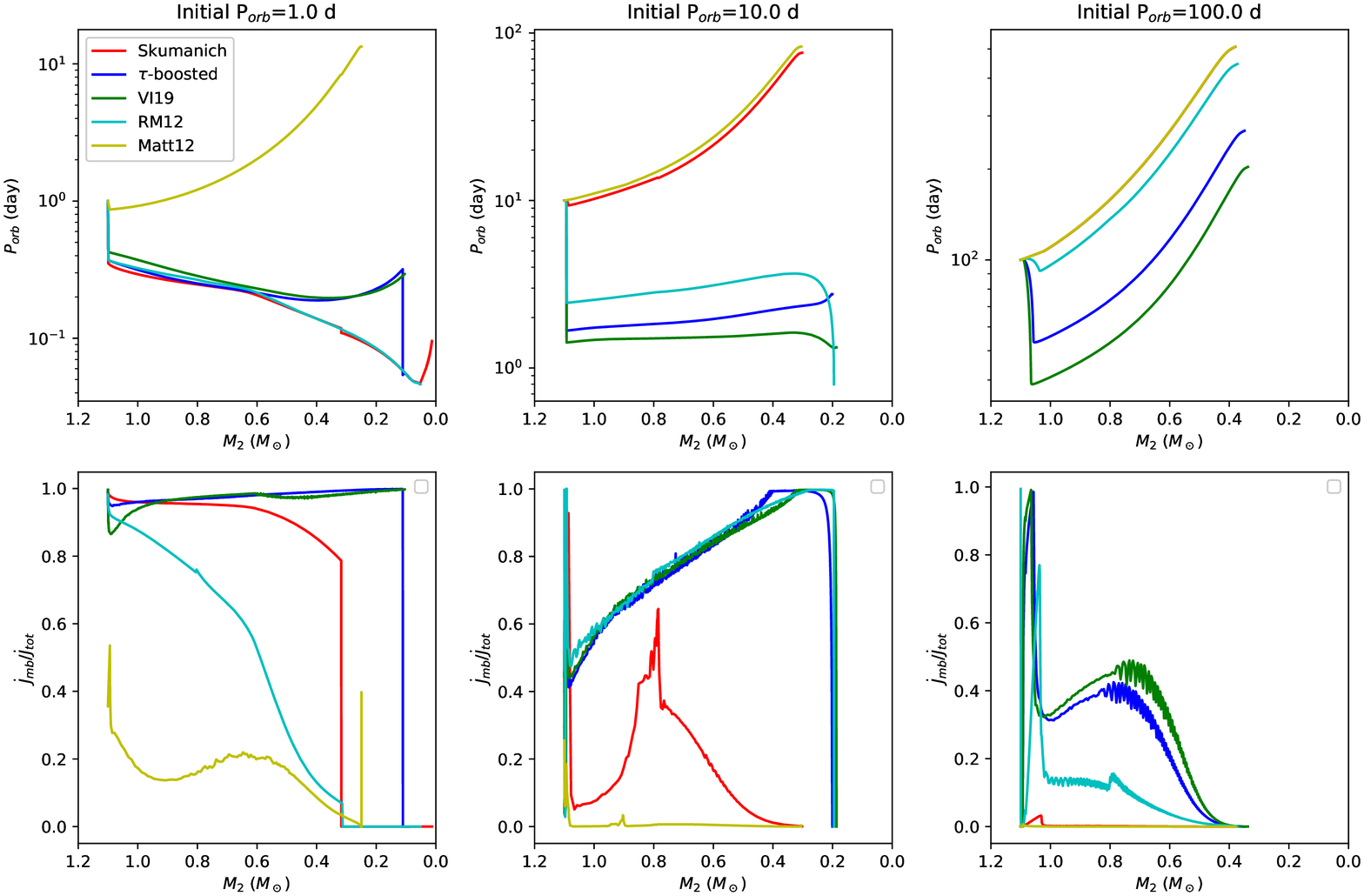}}
%\plotone{eps/FIG1.eps}
\caption{Evolution of a LMXB by using different MB formulae. The red, blue, green, cyan and yellow curves are for the Skumanich, $\tau$-boosted, VI19, RM12, and Matt12 MB laws, respectively. In the left, middle, and right panels, the initial orbital period is taken to be 1, 10, and 100 d, respectively. The top and bottom panels show the evolution of $P_{\rm orb}$ and the rate of AML due to MB divided by the total AML rate as a function of the donor mass, respectively.
    \label{figure1}}

\end{figure}

\clearpage

\begin{sidewaysfigure}

\centering

\subfigure{\includegraphics[scale=0.50]{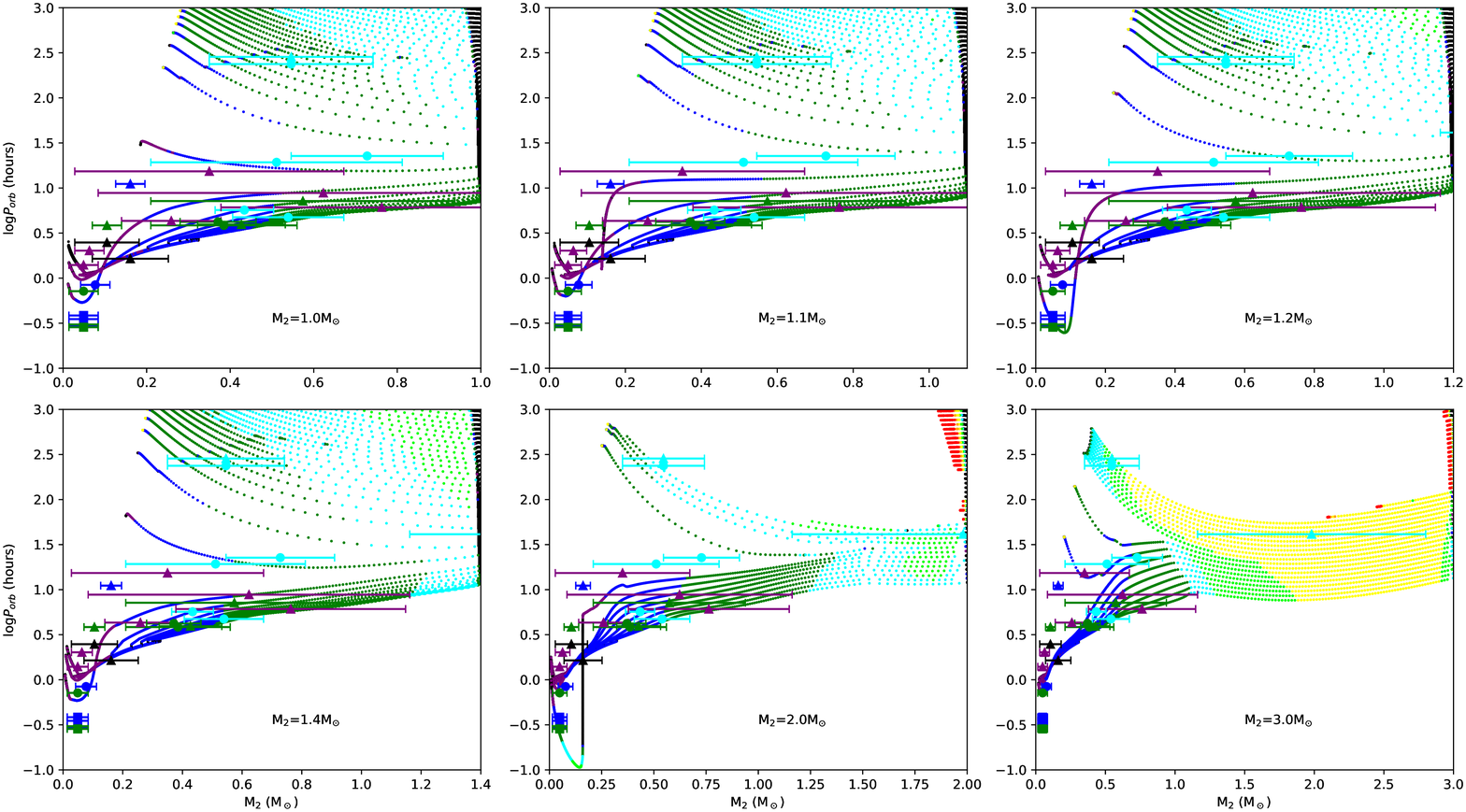}}
\subfigure{\includegraphics[scale=0.50]{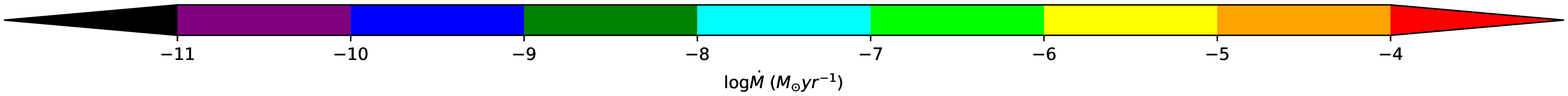}}

\caption{I/LMXB evolution under the Skumanich MB law. Different colors denote the magnitude of the MT rates. The symbols with errorbars represent observed systems, with circles, squares and triangles corresponding to persistent, GC, and transient LMXBs, respectively.
    \label{figure2}}
\end{sidewaysfigure}

\begin{sidewaysfigure}

\centering

\subfigure{\includegraphics[scale=0.5]{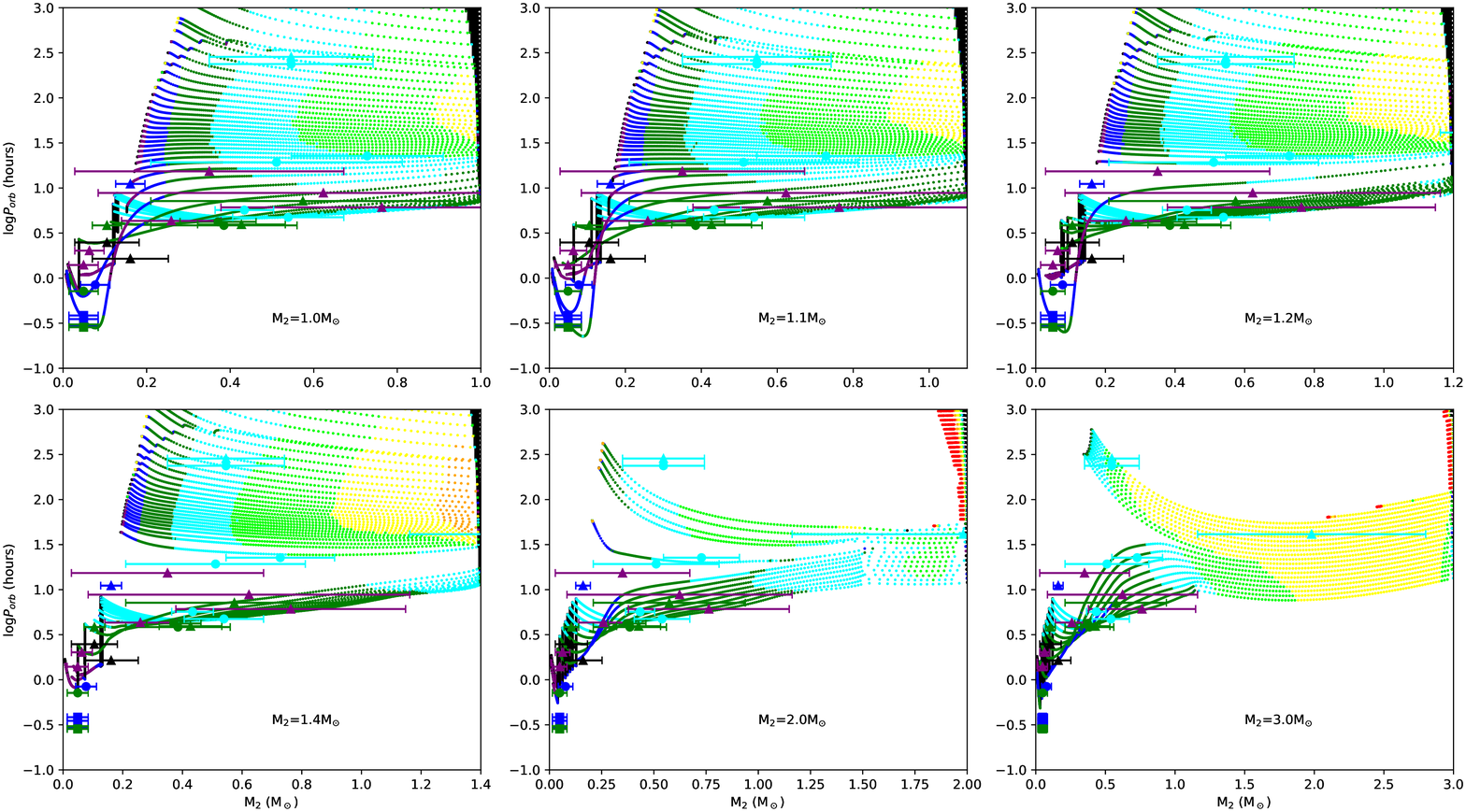}}
\subfigure{\includegraphics[scale=0.5]{colorbar.eps}}

\caption{Same as Fig.~2 but under the $\tau$-boosted MB law.
   \label{figure3}}

\end{sidewaysfigure}

\begin{sidewaysfigure}

\centering

\subfigure{\includegraphics[scale=0.5]{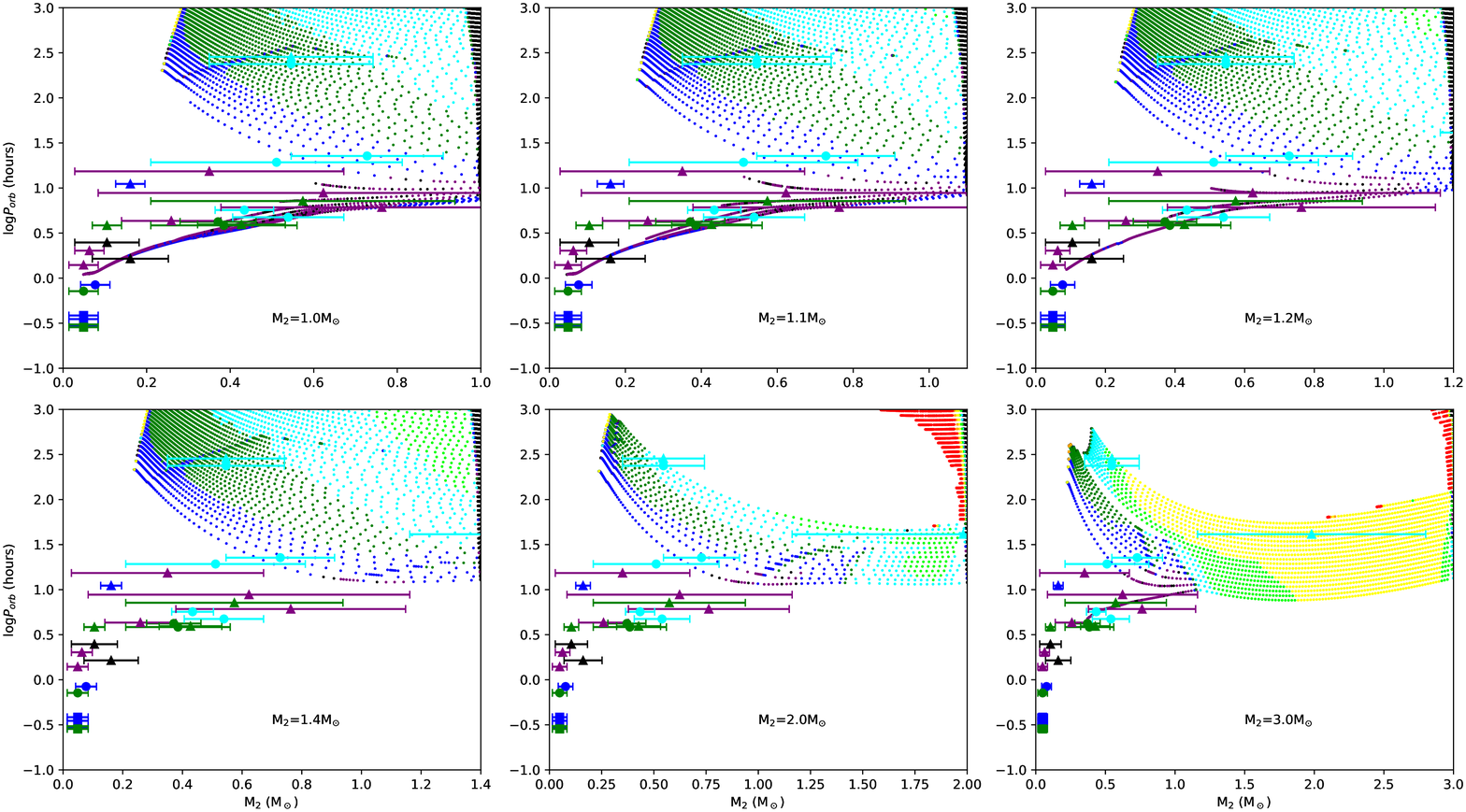}}
\subfigure{\includegraphics[scale=0.5]{colorbar.eps}}

\caption{Same as Fig.~2 but under the Matt12 MB law.
   \label{figure4}}

\end{sidewaysfigure}

\begin{sidewaysfigure}

\centering

\subfigure{\includegraphics[scale=0.5]{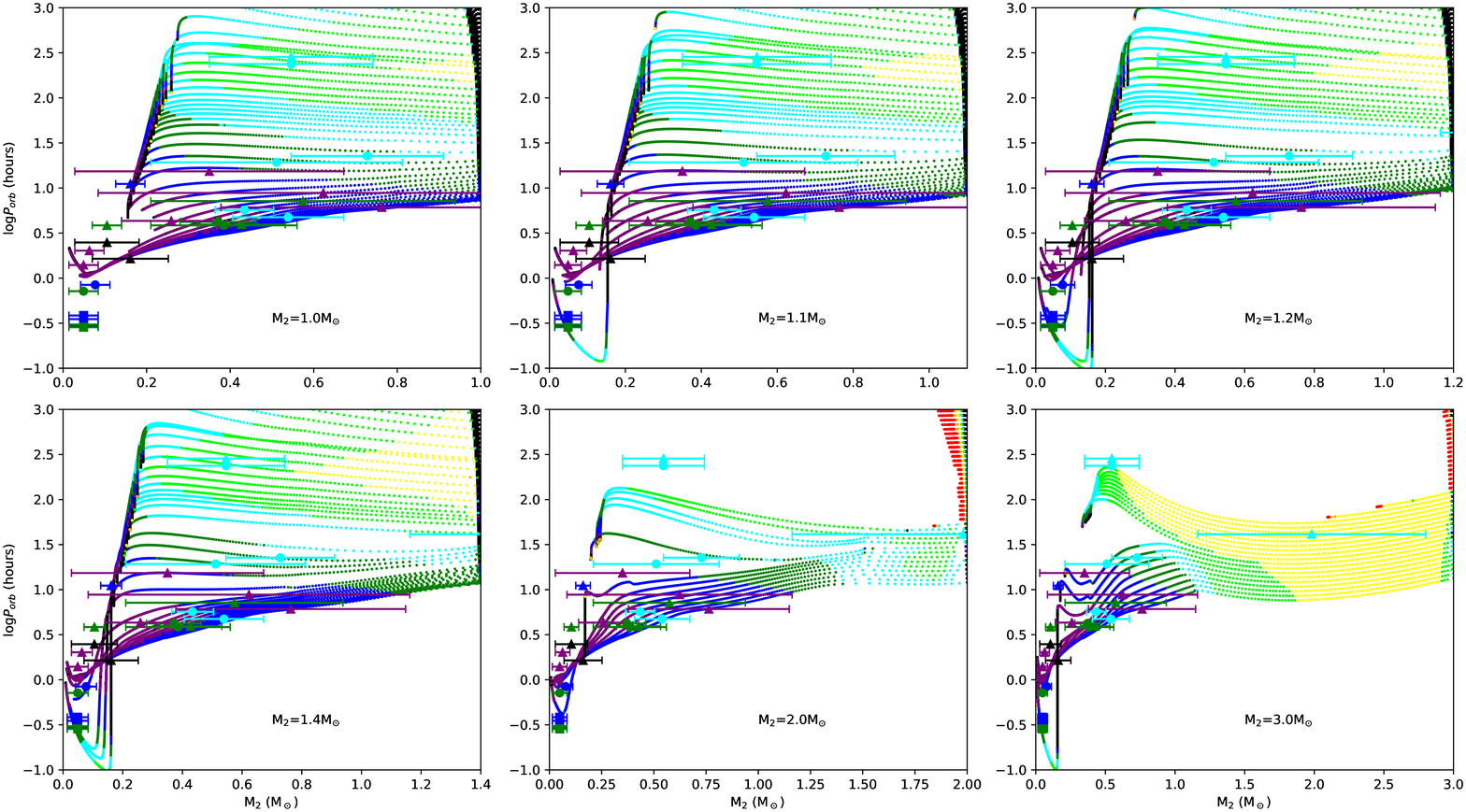}}
\subfigure{\includegraphics[scale=0.5]{colorbar.eps}}
\caption{Same as Fig.~2 but under the RM12 MB law.
   \label{figure5}}

\end{sidewaysfigure}

\begin{sidewaysfigure}

\centering

\subfigure{\includegraphics[scale=0.5]{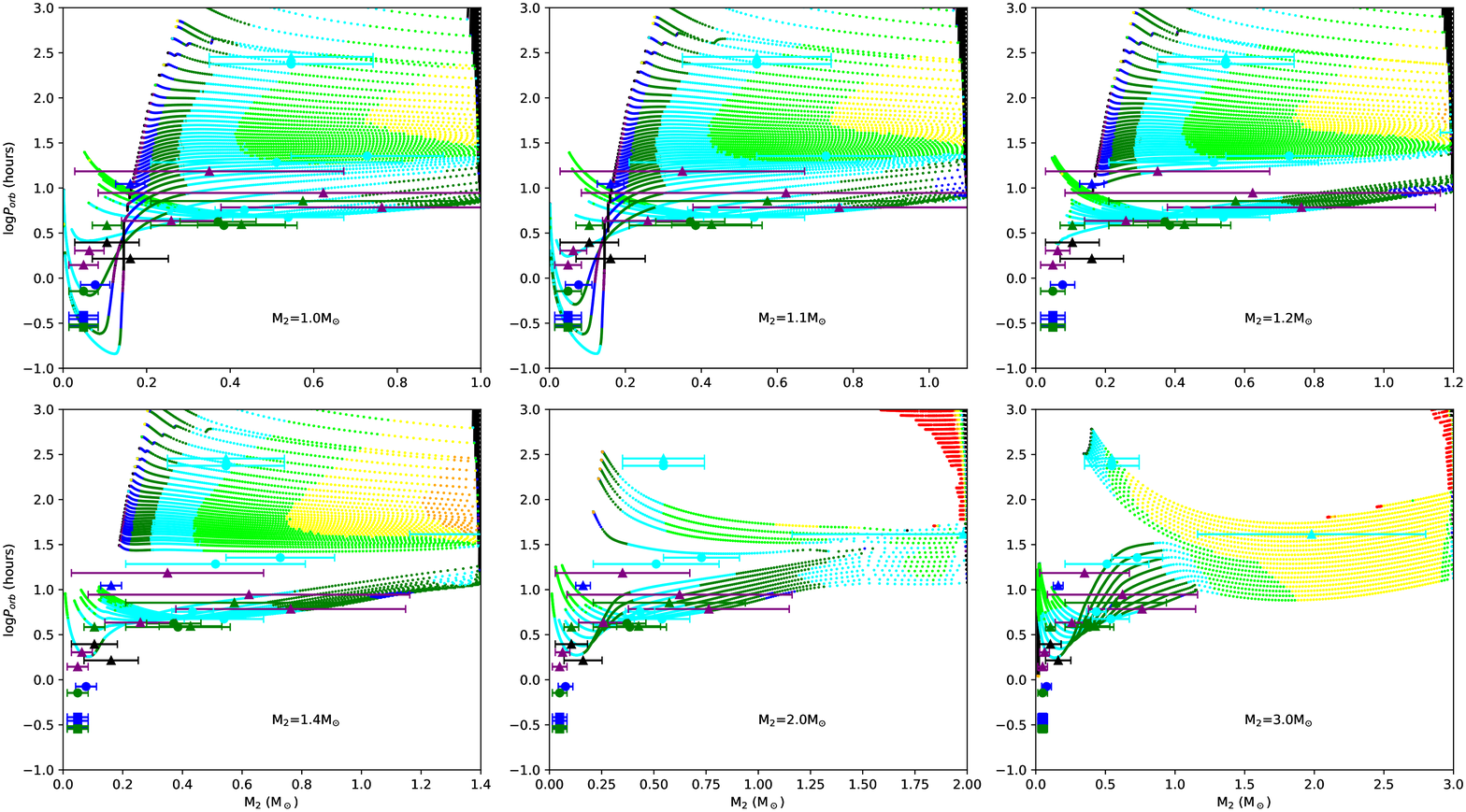}}
\subfigure{\includegraphics[scale=0.5]{colorbar.eps}}
\caption{Same as Fig.~2 but under the VI19 MB law.
   \label{figure6}}

\end{sidewaysfigure}

\begin{figure}

\centering

\subfigure{\includegraphics[scale=0.25]{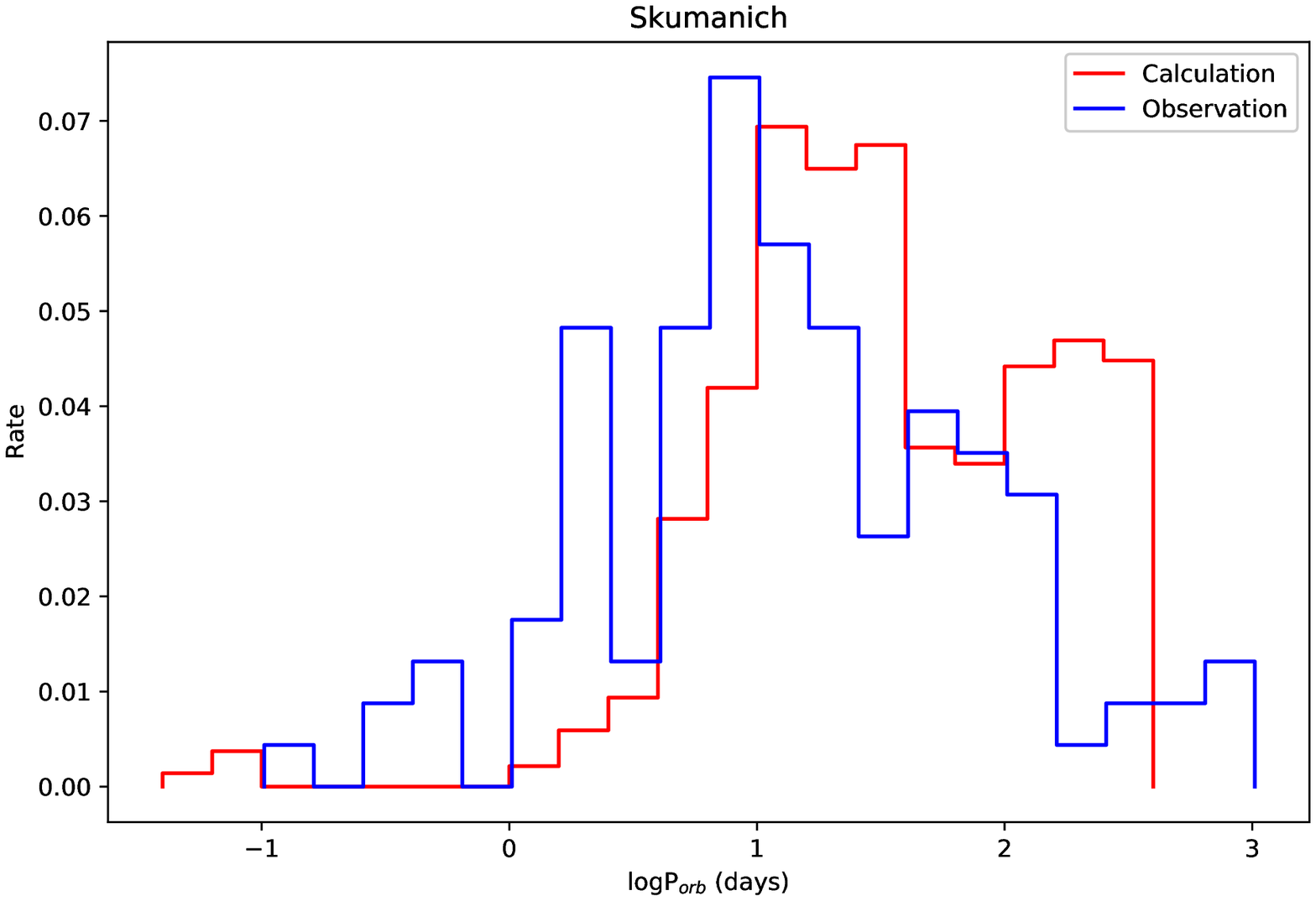}}
\subfigure{\includegraphics[scale=0.25]{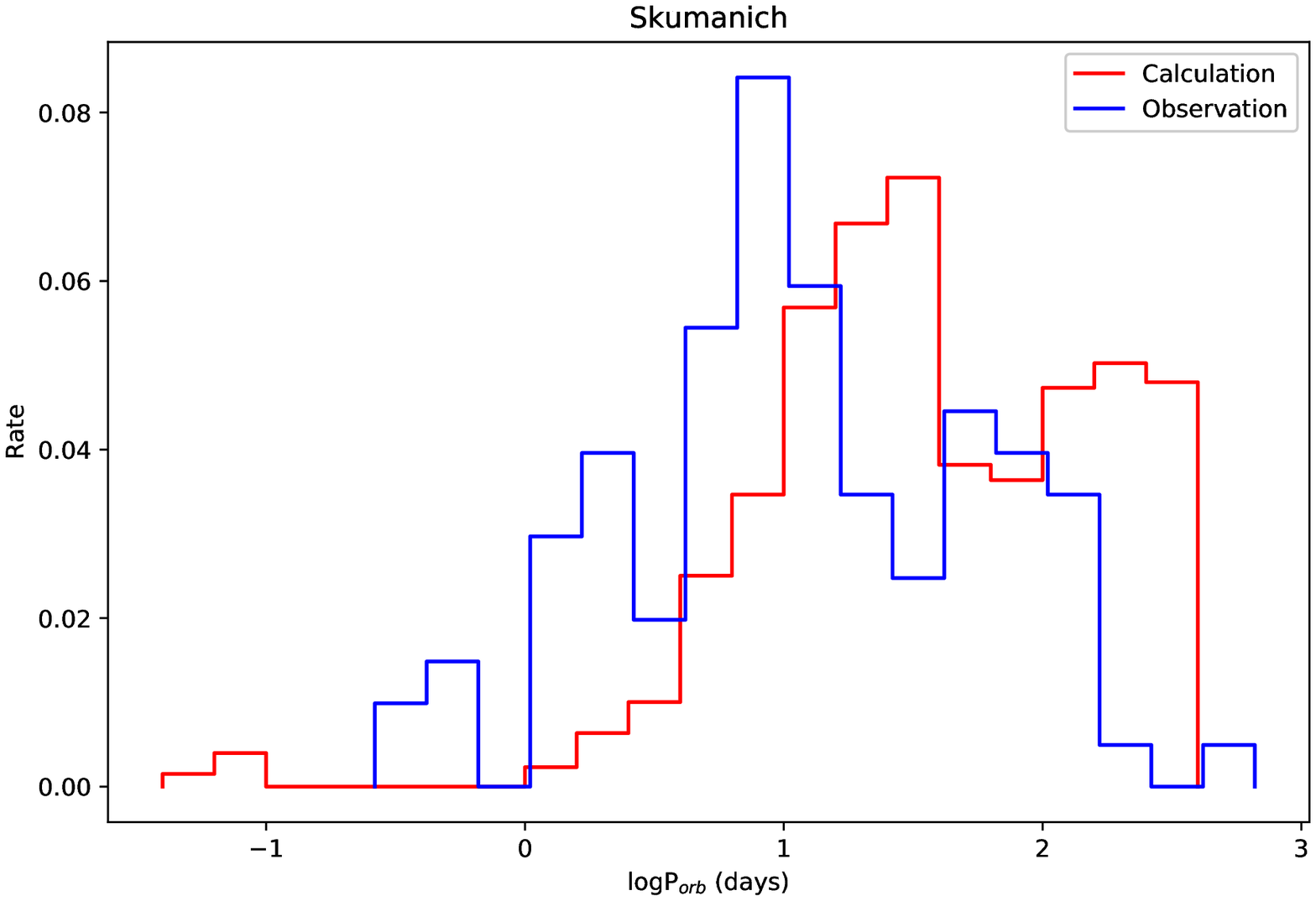}}
\subfigure{\includegraphics[scale=0.25]{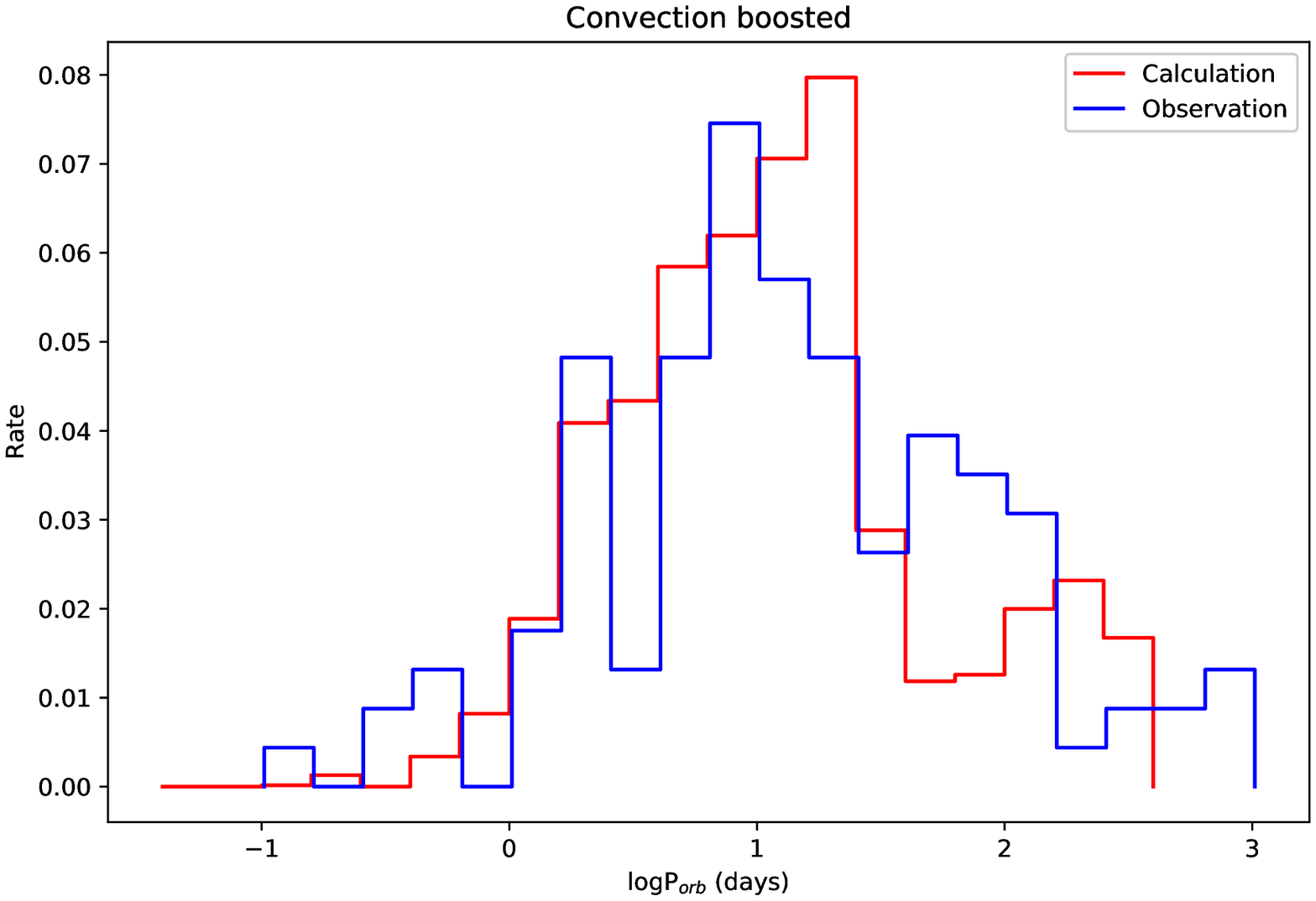}}
\subfigure{\includegraphics[scale=0.25]{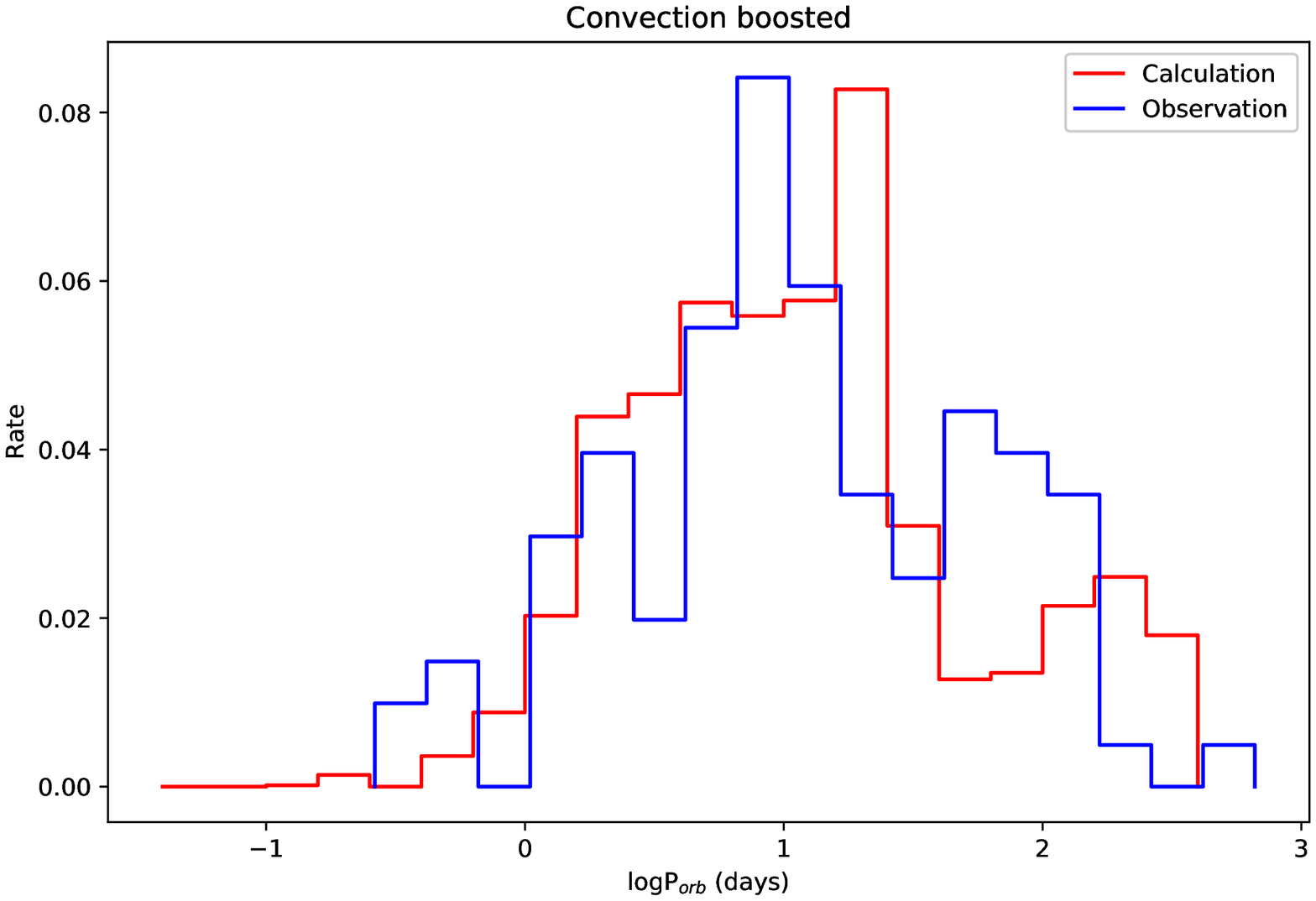}}
\subfigure{\includegraphics[scale=0.25]{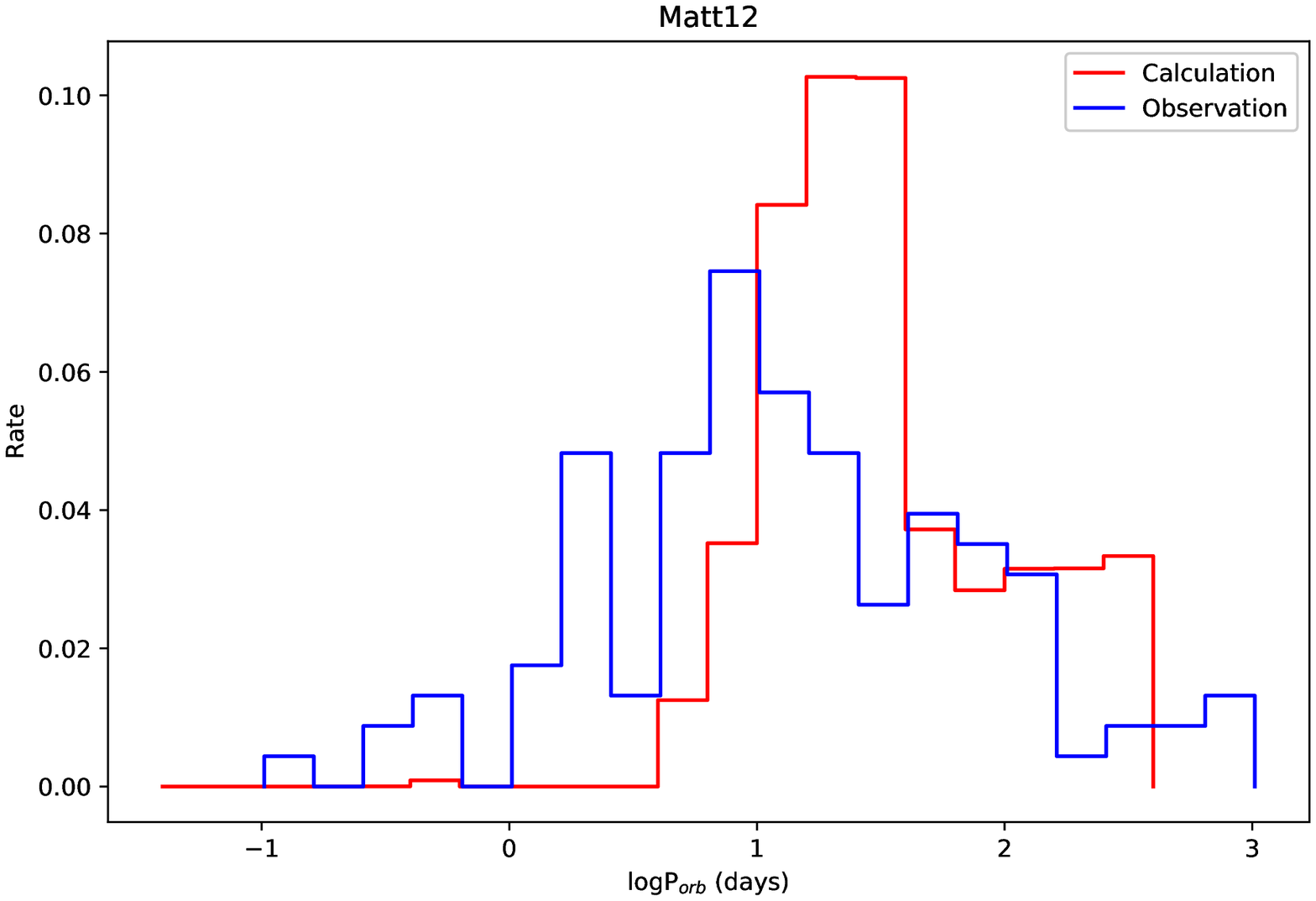}}
\subfigure{\includegraphics[scale=0.25]{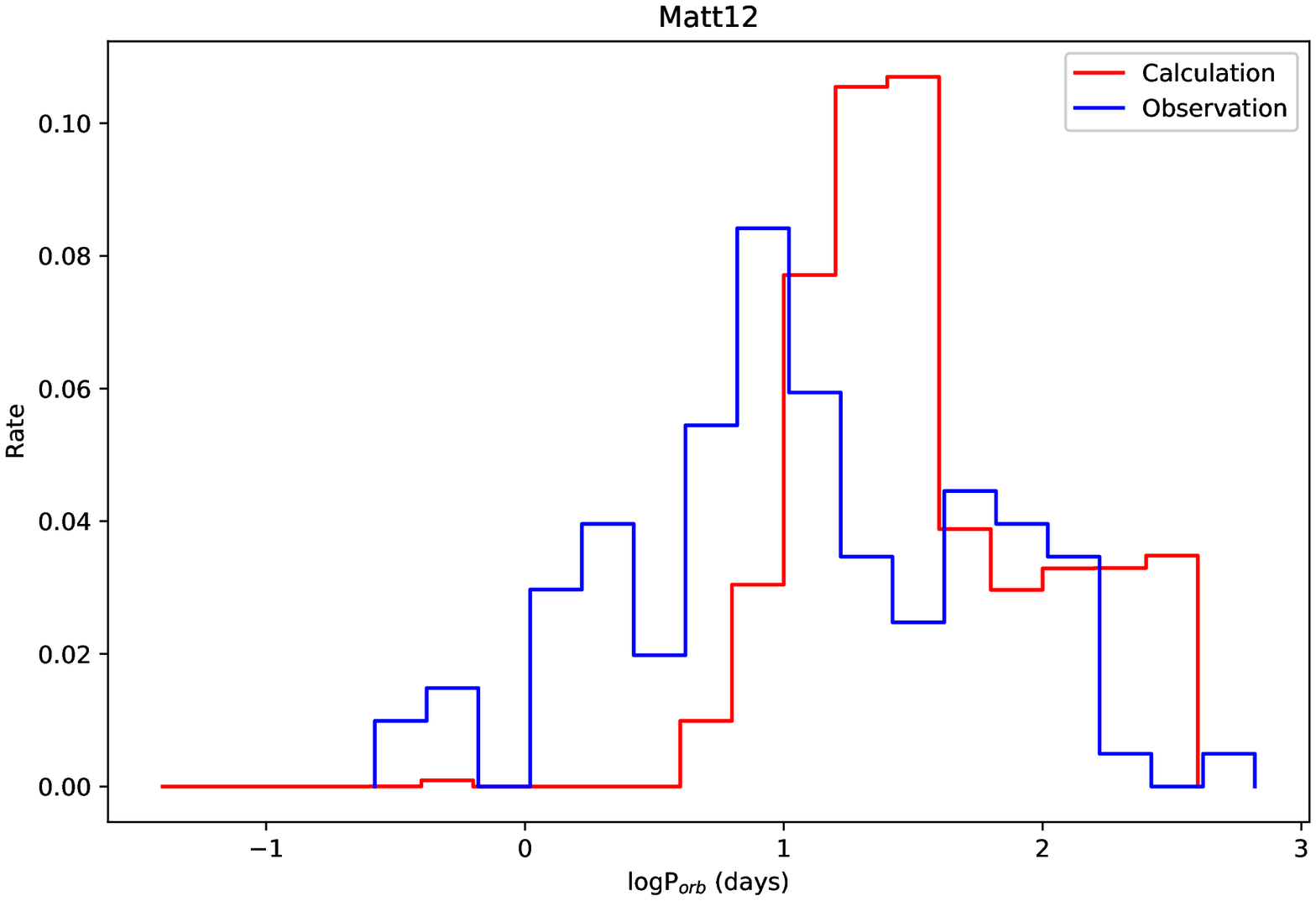}}
\subfigure{\includegraphics[scale=0.25]{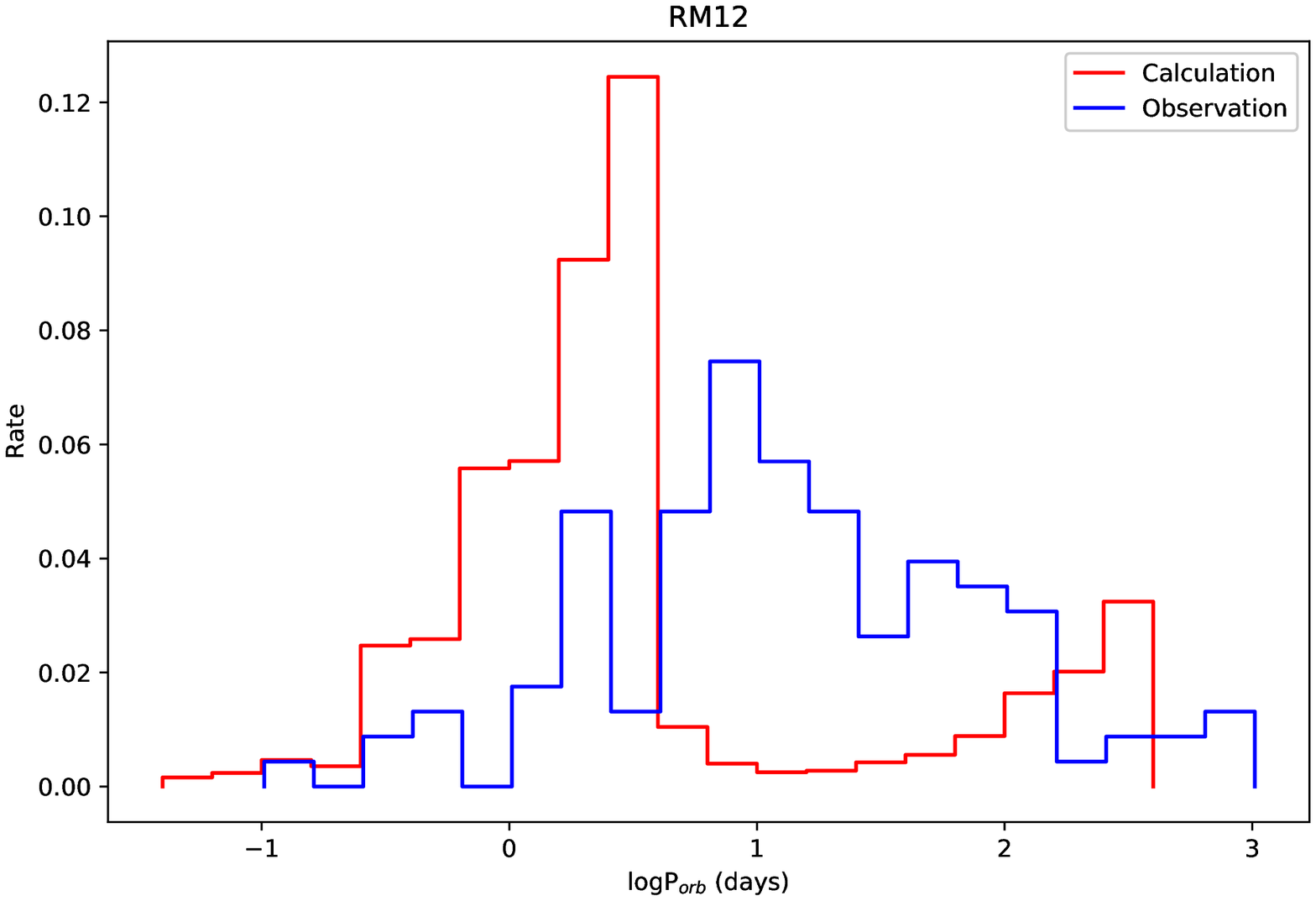}}
\subfigure{\includegraphics[scale=0.25]{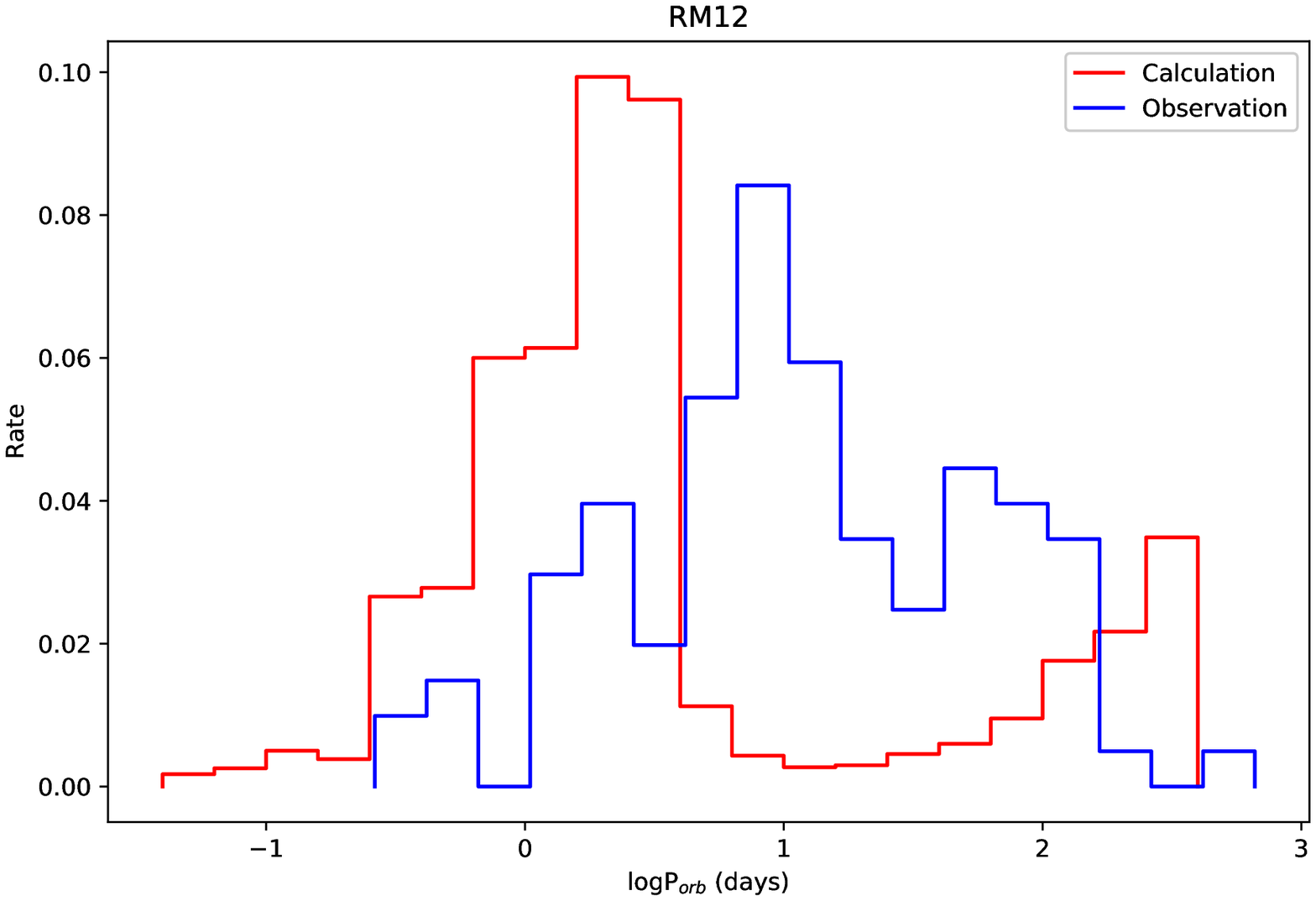}}
\subfigure{\includegraphics[scale=0.25]{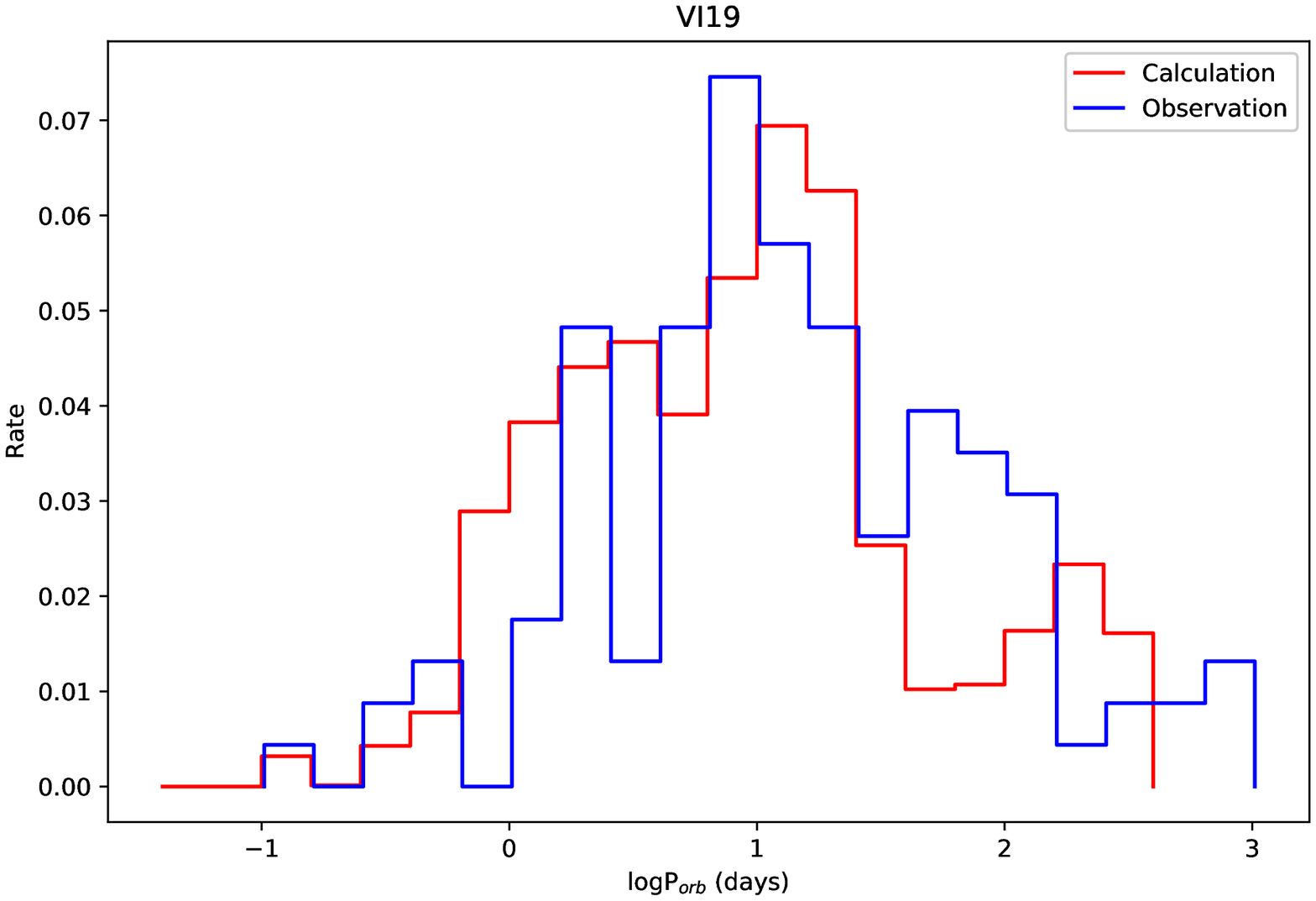}}
\subfigure{\includegraphics[scale=0.25]{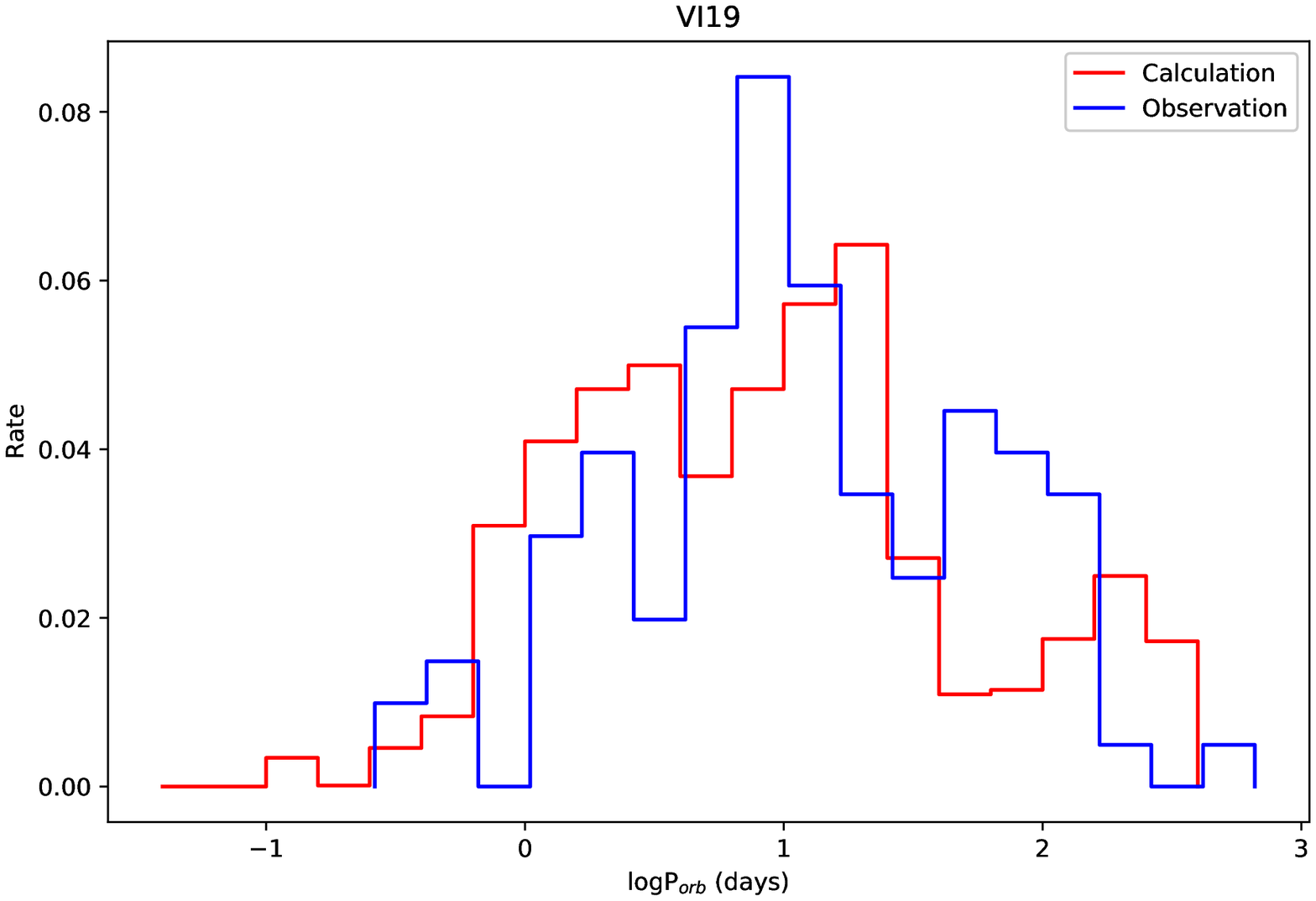}}

\caption{Comparison of the calculated orbital period distributions of binary pulsars with observations, which are shown with the red and blue curves, respectively. The left panels show the calculated distributions of binary pulsars with accreted mass $\Delta M_{\rm NS}>0$ and the observed distribution of binary pulsars with spin periods $P_{\rm s} \leq 1$ s. The right panels show the calculated distributions of binary pulsars with accreted mass $\Delta M_{\rm NS}>0.05M_{\odot}$ and the observed distribution of binary pulsars with the spin periods $P_{\rm s} \leq 30$ ms.
\label{figure7}}

\end{figure}

\clearpage

\begin{figure}

\centering

\centerline{\includegraphics[scale=0.5]{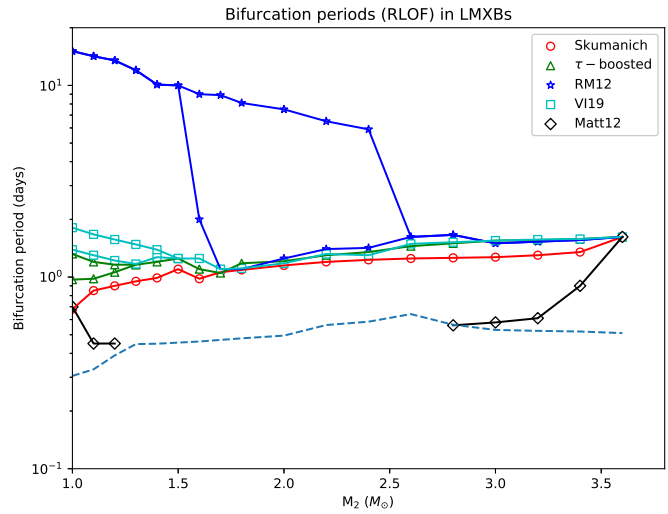}}

\caption{Bifurcation periods as a function of the secondary mass in an LMXB under the five kinds of MB laws. The dotted line shows the minimum initial period $P_{\rm ZAMS}$ that corresponds to a Roche-lobe filling ZAMS secondary star.
    \label{figure1}}

\end{figure}

\clearpage

\begin{sidewaysfigure}

\centering

\subfigure{\includegraphics[scale=0.32]{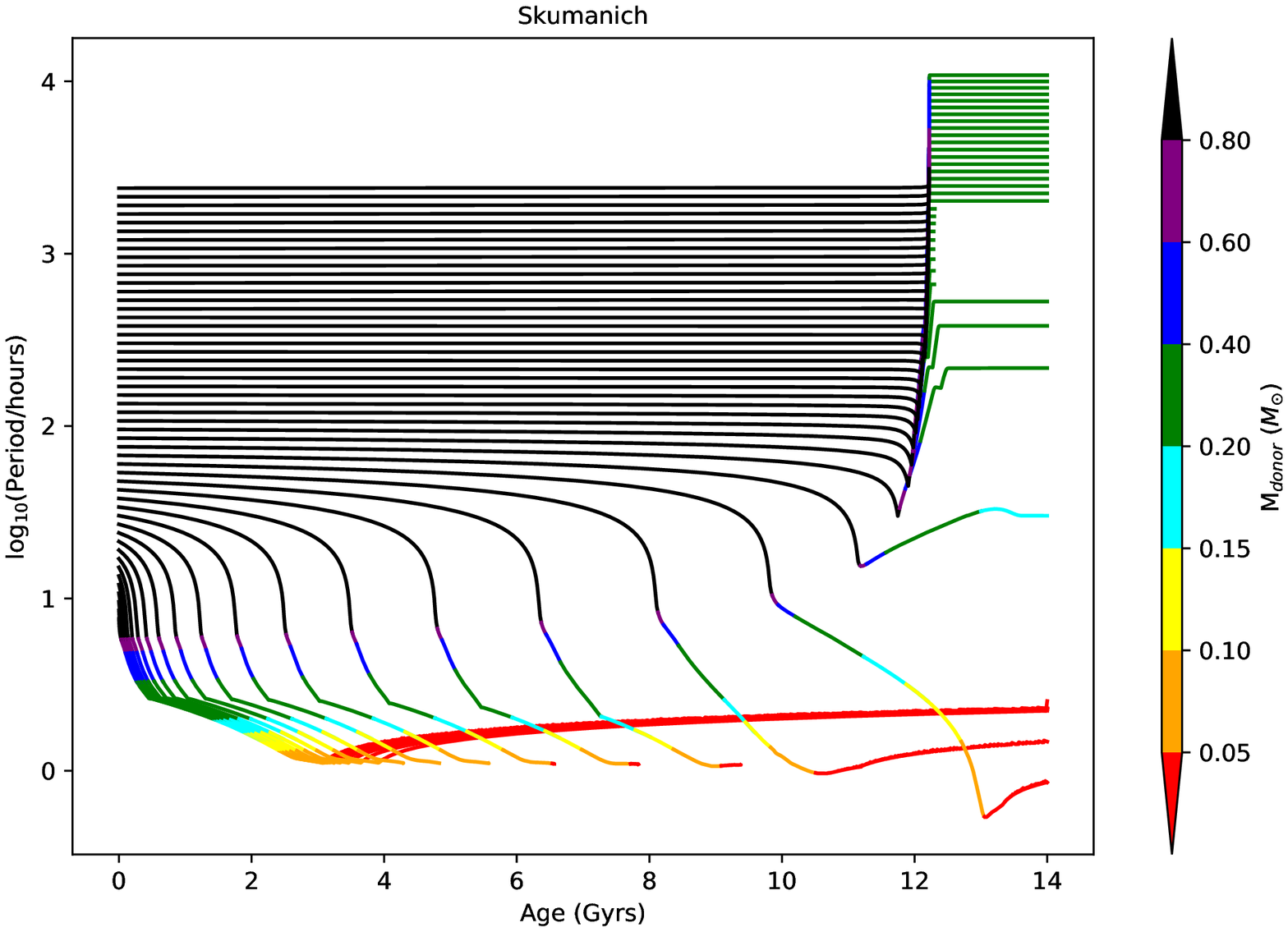}}
\subfigure{\includegraphics[scale=0.32]{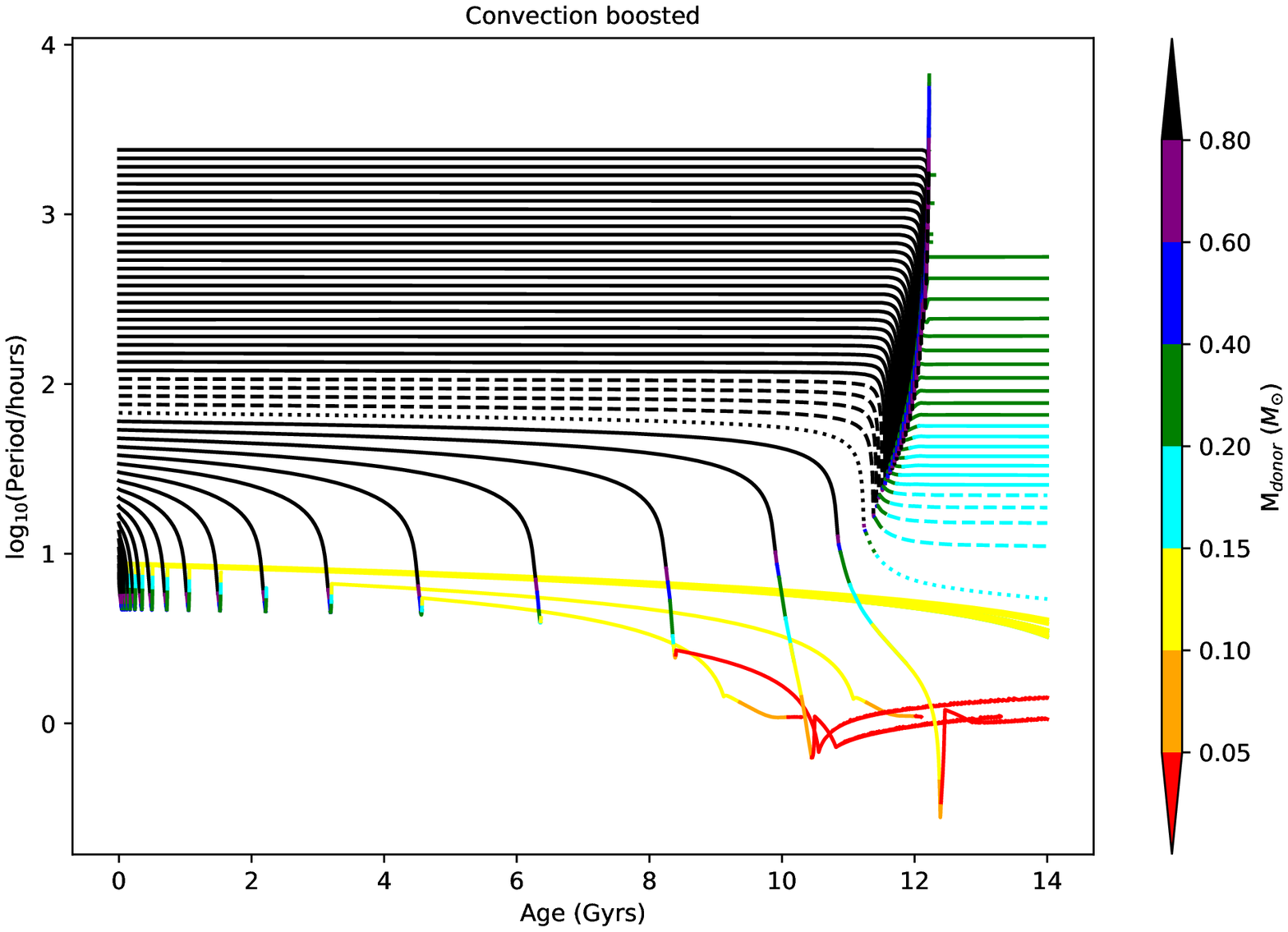}}
\subfigure{\includegraphics[scale=0.32]{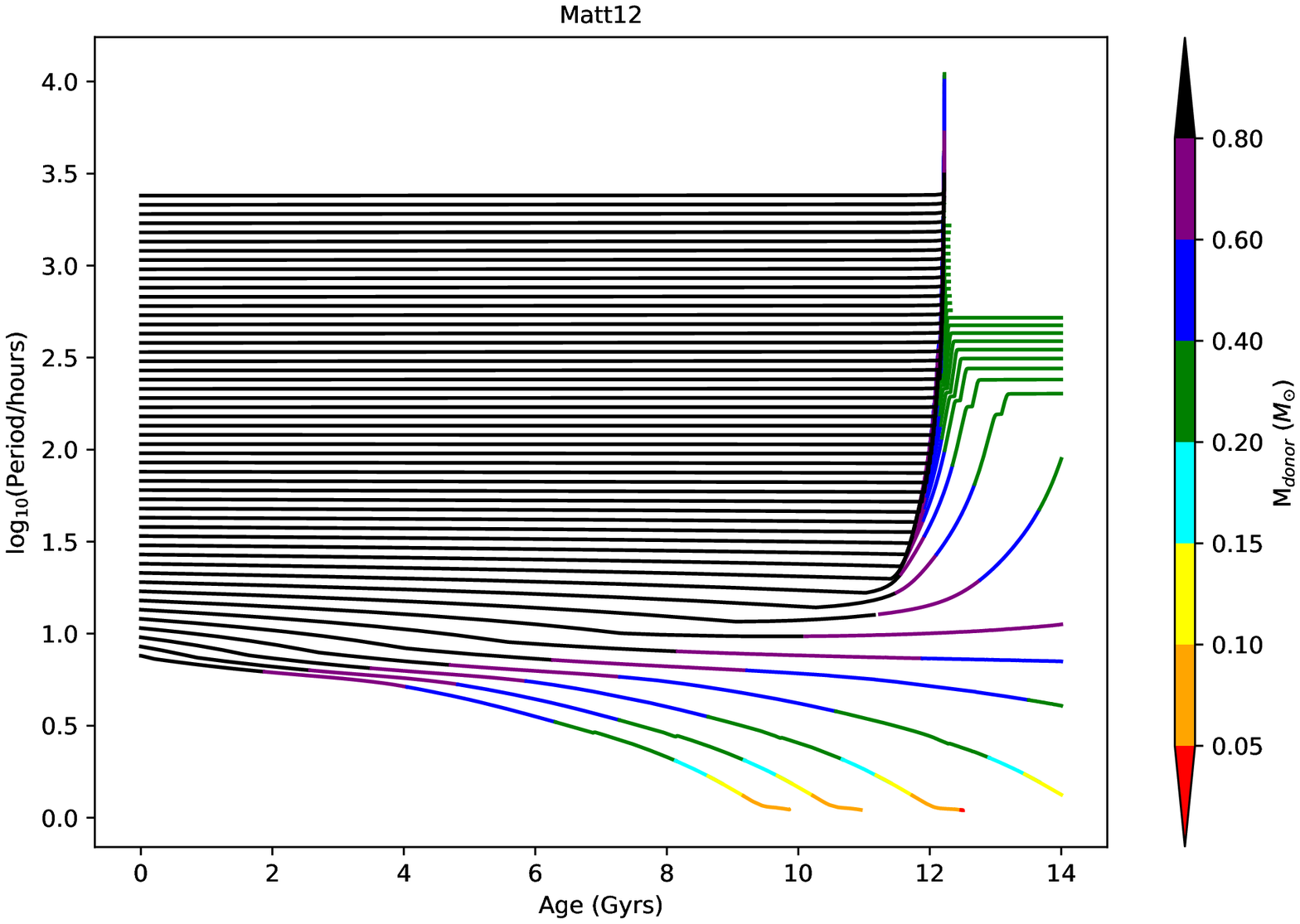}}
\subfigure{\includegraphics[scale=0.32]{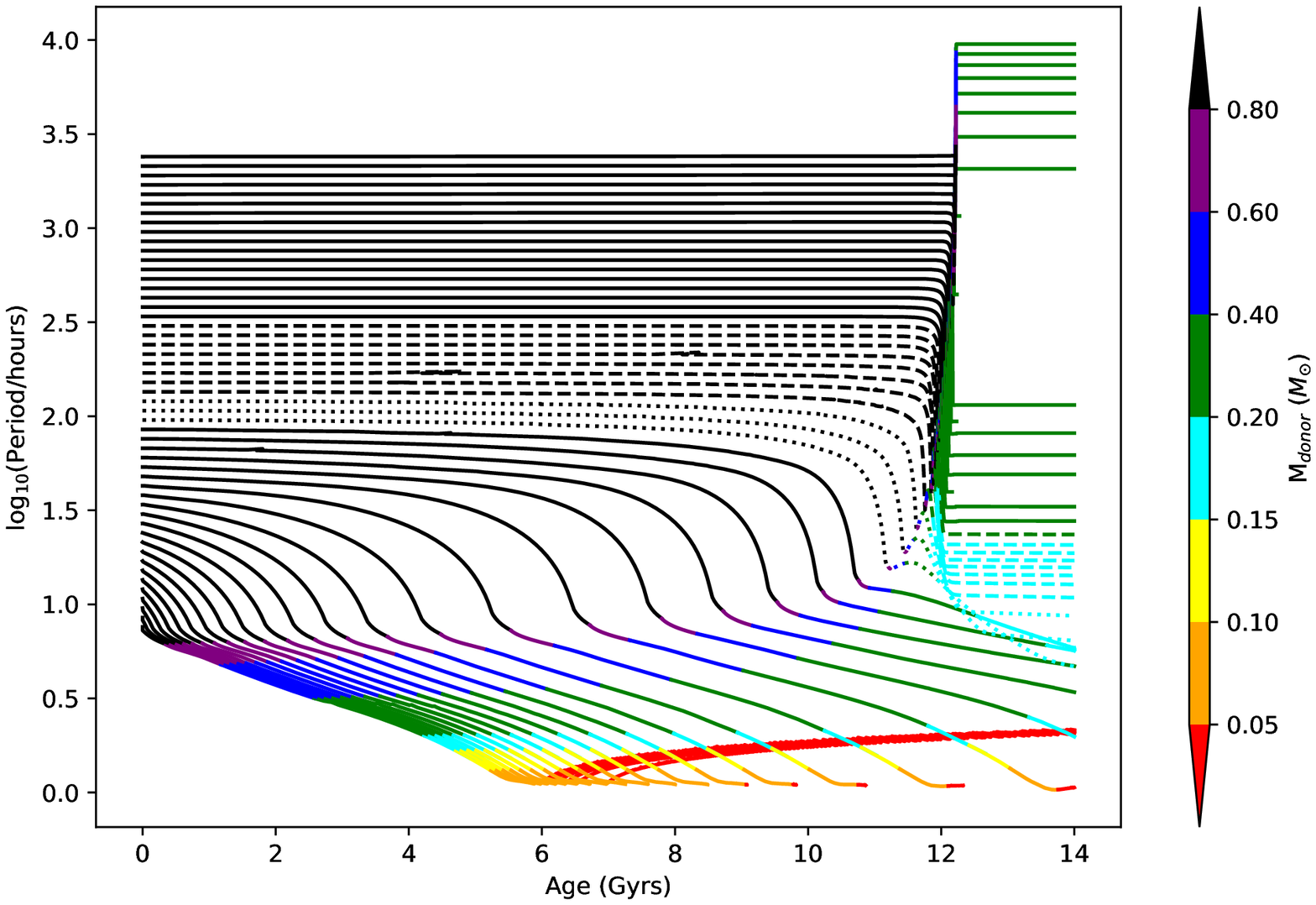}}
\subfigure{\includegraphics[scale=0.32]{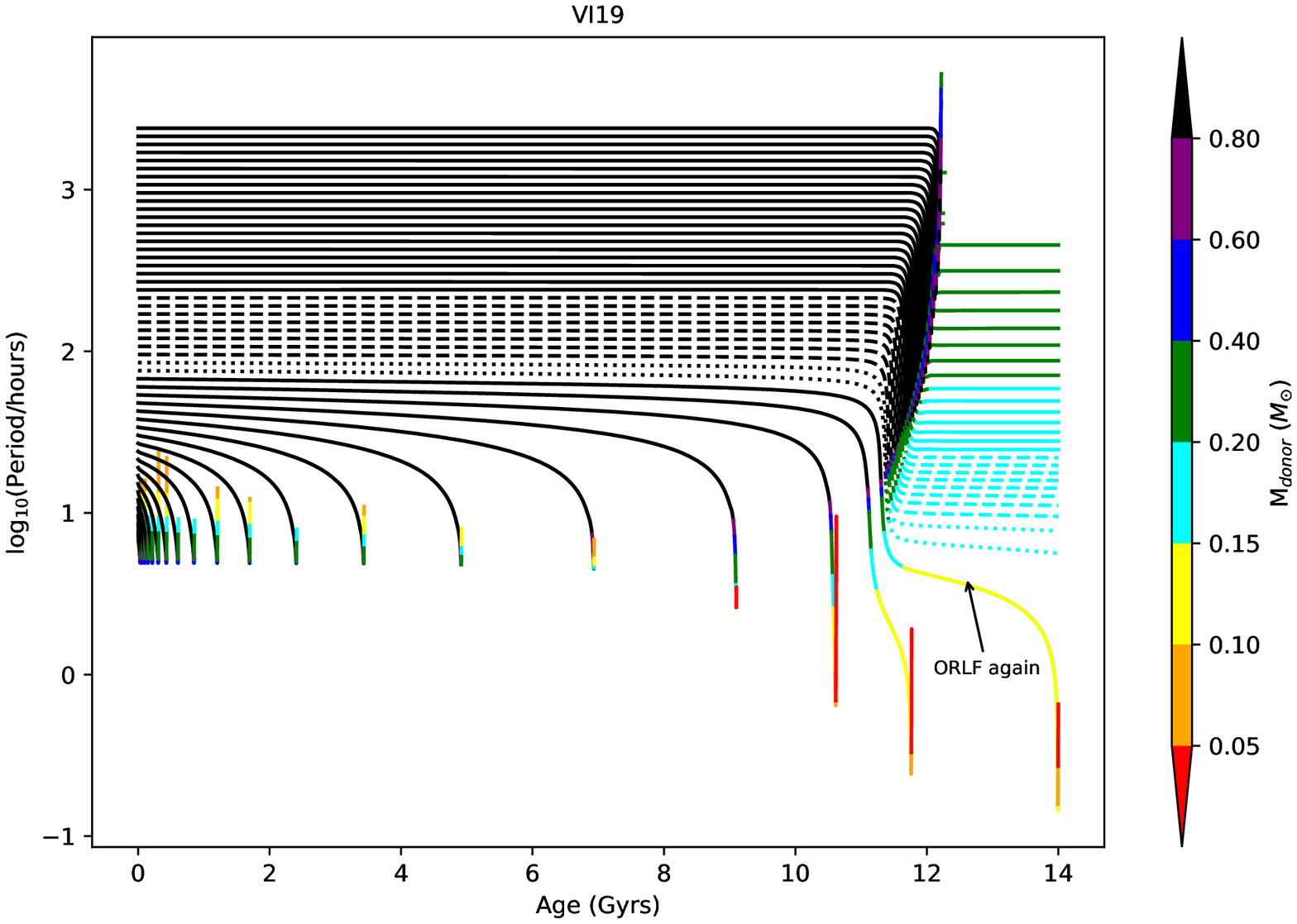}}

\caption{Orbital period evolution for LMXBs with a donor of mass of $1.0 M_{\odot}$ in the five MB models. The dashed and dotted curves denote forming BMSPs with orbital periods $P_{\rm orb}=9-24$ h and $P_{\rm orb}=2-9$ h, respectively. Different colors denote the donor mass.
    \label{figure8}}

\end{sidewaysfigure}

\clearpage

\begin{table}
%\tiny
%\scriptsize
\footnotesize
%\small
%\normalsize
%\large
%\huge
\begin{center}
\caption{Calculated Results of the Binary Evolution with $M_{\rm 1,i}=1.3M_{\odot}$ and $M_{\rm 2, i}=1.1M_{\odot}$ by using different MB laws.}
\begin{tabular}{c|cccc|cccc|cccc}
  \toprule[1pt]
   & \multicolumn{4}{c}{$P_{\rm orb,i}$=1.0 d} & \multicolumn{4}{c}{$P_{\rm orb,i}$=10.0 d} & \multicolumn{4}{c}{$P_{\rm orb,i}$=100.0 d} \\
  \hline
   MB  & $M_{\rm 1,f}$     & $M_{\rm 2,f}$     & $P_{\rm orb,f}$ & $H_{\rm f}$  & $M_{\rm 1,f}$     & $M_{\rm 2,f}$     & $P_{\rm orb,f}$ & $H_{f}$ & $M_{\rm 1,f}$     &  $M_{\rm 2,f}$    & $P_{\rm orb,f}$ & $H_{\rm f}$  \\
   model& ($M_{\odot}$) & ($M_{\odot}$) & (days)          &         (\%)       & ($M_{\odot}$) & ($M_{\odot}$) &      (days)     &          (\%)     & ($M_{\odot}$) & ($M_{\odot}$) & (days)          & (\%)  \\
  \hline
  Skumanich    & 2.261 & 0.013 & 0.096 & 68.2 & 1.879 & 0.302 & 76.264 & 0.3 & 1.403 & 0.380 & 508.447 & 0.06 \\
  $\tau$-boosted     & 2.185 & 0.088 & 0.051 & 62.0 & 1.578 & 0.201 & 2.758 & 0.5 & 1.467 & 0.349 & 267.892 & 0.2 \\
  Matt12       & 2.034 & 0.250 & 13.367 & 0.2 & 1.892 & 0.305 & 83.024 & 0.4 & 1.402 & 0.380 & 509.214 & 0.1  \\
  RM12         & 2.230 & 0.051 & 0.046 & 65.2 & 1.607 & 0.195 & 0.8 & 0.004 & 1.424 & 0.373 & 445.8 & 0.001 \\
  VI19         & 1.979 & 0.104 & 0.294 & 46.7 & 1.492 & 0.187 & 1.327 & 0.5 & 1.516 & 0.338 & 203.295 & 0.3 \\
   \bottomrule[1pt]

\end{tabular}
\end{center}
Note: $M_{\rm 1,f}$, $M_{\rm 2,f}$, $P_{\rm orb,f}$ and $H_{\rm f}$ are the masses of the NS and the companion star, the orbital period and the H abundance of companion star at the end of the mass transfer, respectively.
\end{table}

\clearpage

\begin{table}
%\tiny
%\scriptsize
\footnotesize
%\small
%\normalsize
%\large
%\huge
\begin{center}
\caption{Comparison of the effectiveness of different MB models in reproducing the properties of persistent (upper) and transient (lower) LMXBs}
\begin{tabular}{l|c|c|c|c|c}
  \toprule[1pt]

Source &Skumanich &$\tau$-boosted &Matt12 &RM12 &VI19  \\
  \hline

4U 0513-40 & $\blacktriangle $ & $\blacktriangle \blacktriangle$ & $\triangle$ & $\blacktriangle \blacktriangle \blacktriangle$ & $\blacktriangle \blacktriangle$  \\
 2S 0918-549 & $\blacktriangle $ & $\blacktriangle \blacktriangle $ & $\triangle$ & $\blacktriangle \blacktriangle \blacktriangle$ & $\blacktriangle \blacktriangle$  \\
 4U 1543-624 & $\blacktriangle $ & $\blacktriangle \blacktriangle$ & $\triangle$ & $\blacktriangle \blacktriangle \blacktriangle$ & $\blacktriangle \blacktriangle$  \\
 4U 1850-087 & $\blacktriangle $ & $\blacktriangle \blacktriangle$ & $\triangle$ & $\blacktriangle \blacktriangle \blacktriangle$ & $\blacktriangle \blacktriangle$  \\
 M15 X-2 & $\blacktriangle $ & $\blacktriangle \blacktriangle$ & $\triangle$ & $\blacktriangle \blacktriangle \blacktriangle$ & $\blacktriangle \blacktriangle$  \\
 4U 1626-67 & $\blacktriangle \blacktriangle $ & $\blacktriangle \blacktriangle \blacktriangle$ & $\triangle$ & $\blacktriangle$ & $\blacktriangle \blacktriangle$  \\
 4U 1916-053 & $\blacktriangle \blacktriangle$ & $\blacktriangle$ & $\triangle$ & $\triangle$ & $\blacktriangle \blacktriangle$  \\
 4U 1636-536 & $\blacktriangle \blacktriangle \blacktriangle$ & $\blacktriangle \blacktriangle \blacktriangle$ & $\triangle$ & $\triangle$ & $\blacktriangle \blacktriangle \blacktriangle$  \\
 GX 9+9 & $\triangle$ & $\blacktriangle \blacktriangle \blacktriangle$ & $\triangle$ & $\triangle$ & $\blacktriangle \blacktriangle \blacktriangle$  \\
 4U 1735-444 & $\triangle$ & $\blacktriangle \blacktriangle \blacktriangle$ & $\triangle$ & $\triangle$ & $\blacktriangle \blacktriangle \blacktriangle$  \\
 2A 1822-371 & $\triangle$ & $\triangle$ & $\triangle$ & $\triangle$ & $\blacktriangle \blacktriangle \blacktriangle$  \\
 Sco X-1 & $\triangle$ & $\blacktriangle \blacktriangle \blacktriangle$ & $\triangle$ & $\triangle$ & $\blacktriangle \blacktriangle \blacktriangle$ \\
 GX 349+2 & $\blacktriangle$ & $\blacktriangle \blacktriangle \blacktriangle$ & $\blacktriangle$ & $\blacktriangle$ & $\blacktriangle \blacktriangle$ \\
 Cyg X-2 & $\blacktriangle \blacktriangle \blacktriangle$ & $\blacktriangle \blacktriangle \blacktriangle$ & $\blacktriangle \blacktriangle \blacktriangle$ & $\triangle$ & $\blacktriangle \blacktriangle \blacktriangle$  \\
%
%    \bottomrule[1pt]
%
%\end{tabular}
%\end{center}
%Formation of persistent LMXBs under different MB laws. The symbols with $\triangle$, $\blacktriangle$, $\blacktriangle \blacktriangle$ and $\blacktriangle \blacktriangle \blacktriangle$ represent that 0, 1$\sim$2, 3$\sim$5 and larger than 5 models to reproduce observed systems, respectively.
%
%\end{table}
%
%\clearpage
%
%\begin{table}
%%\tiny
%%\scriptsize
%\footnotesize
%%\small
%%\normalsize
%%\large
%%\huge
%\begin{center}
%\caption{Transient LMXBs}
%\begin{tabular}{c|c|c|c|c|c}
%  \toprule[1pt]
%System name &Skumanich &$\tau$-boosted &Matt12 &RM12 &VI19  \\
  \hline

 HETE J1900.1-2455 & $\blacktriangle \blacktriangle \blacktriangle$ & $\blacktriangle \blacktriangle \blacktriangle$ & $\triangle$ & $\blacktriangle \blacktriangle \blacktriangle$ & $\triangle$  \\
 1A 1744-361 & $\blacktriangle \blacktriangle \blacktriangle$ & $\blacktriangle \blacktriangle$ & $\blacktriangle \blacktriangle$ & $\blacktriangle \blacktriangle \blacktriangle$ & $\blacktriangle$  \\
 SAX J1808-3658 & $\triangle$ & $\triangle$ & $\triangle$ & $\triangle$ & $\triangle$  \\
 IGR 00291+5394 & $\blacktriangle \blacktriangle \blacktriangle$ & $\blacktriangle \blacktriangle$ & $\triangle$ & $\blacktriangle \blacktriangle \blacktriangle$ & $\blacktriangle$  \\
 EXO 0748-678 & $\triangle$ & $\blacktriangle \blacktriangle$ & $\triangle$ & $\triangle$ & $\triangle$  \\
 4U 1254-69 & $\blacktriangle \blacktriangle \blacktriangle$ & $\blacktriangle \blacktriangle \blacktriangle$ & $\triangle$ & $\triangle$ & $\blacktriangle \blacktriangle \blacktriangle$  \\
 XTE J1814-338 & $\blacktriangle \blacktriangle \blacktriangle$ & $\blacktriangle \blacktriangle$ & $\blacktriangle$ & $\blacktriangle \blacktriangle \blacktriangle$ & $\blacktriangle$  \\
 XTE J2123-058 & $\triangle$ & $\triangle$ & $\blacktriangle \blacktriangle \blacktriangle$ & $\blacktriangle \blacktriangle \blacktriangle$ & $\triangle$  \\
 X 1658-298 & $\blacktriangle \blacktriangle \blacktriangle$ & $\blacktriangle \blacktriangle \blacktriangle$ & $\triangle$ & $\triangle$ & $\blacktriangle \blacktriangle \blacktriangle$  \\
 SAX J1748.9-2021 & $\blacktriangle \blacktriangle \blacktriangle$ & $\blacktriangle \blacktriangle$ & $\blacktriangle \blacktriangle \blacktriangle$ & $\blacktriangle \blacktriangle \blacktriangle$ & $\blacktriangle$  \\
 IGR J18245-2452 & $\triangle$ & $\triangle$ & $\triangle$ & $\blacktriangle \blacktriangle \blacktriangle$ & $\blacktriangle \blacktriangle$  \\
 Cen X-4 & $\triangle$ & $\blacktriangle$ & $\blacktriangle$ & $\blacktriangle \blacktriangle \blacktriangle$ & $\blacktriangle $  \\
 Her X-1 & $\blacktriangle \blacktriangle$ & $\blacktriangle \blacktriangle$ & $\blacktriangle \blacktriangle \blacktriangle$ & $\blacktriangle \blacktriangle$ & $\blacktriangle $  \\
 GRO J1744-28 & $\blacktriangle \blacktriangle \blacktriangle$ & $\blacktriangle \blacktriangle \blacktriangle$ & $\blacktriangle \blacktriangle \blacktriangle$ & $\triangle$ & $\blacktriangle \blacktriangle \blacktriangle$  \\

    \bottomrule[1pt]

\end{tabular}
\end{center}
\end{table}

\clearpage

\begin{table}
%\tiny
%\scriptsize
\footnotesize
%\small
%\normalsize
%\large
%\huge
\begin{center}
\caption{Kolmogorov-Smirnov test for the measured and calculated orbital period distributions of binary pulsars.}
\begin{tabular}{c|c|c|c|c|c|c}
  \toprule[1pt]
  $P_s$ & maximum distance & Skumanich & $\tau$-boosted & Matt12 & RM12 & VI19 \\
  \hline
  $\leq$ 1s & $D_{\rm nm}$ & 0.303 & 0.094 & 0.371 & 0.499 & 0.111 \\
  & $D_{\rm \alpha , nm}$ & 0.147 & 0.149 & 0.140 & 0.149 & 0.147 \\
  \hline
  $\leq$ 30ms & $ D_{\rm nm}$ & 0.388 & 0.116 & 0.442 & 0.470 & 0.134 \\
   & $D_{\rm \alpha , nm}$ & 0.155 & 0.156 & 0.148 & 0.156 & 0.155 \\
  \bottomrule[1pt]
\end{tabular}
\end{center}
\end{table}

\end{document}